\documentclass[12pt,twoside]{article}
\usepackage{correl}

\def\TheTitle{Beyond DMFT: Spin Fluctuations, Pseudogaps and Superconductivity}
\def\TheHeading{Beyond DMFT}
\def\TheAuthor{Karsten Held}
\def\TheInstitute{Institute for Solid State Physics, TU Wien}
\def\TheAddress{1040 Wien, Austria}
\def\TheChapter{5}           

\begin{document}
\MakeTitle
\section{Introduction}
\index{dynamical mean-field theory}
With the success of dynamical mean-field theory (DMFT) \cite{Metzner1989,Georges1992a,Pavarini2022} in calculating strongly correlated electron systems, there have been attempts from the very beginning \cite{Schiller1995} to systematically extend DMFT. 
 The aim is here to keep the good description of DMFT for local electronic correlations, but also to capture non-local correlations beyond.

 Indeed, the local DMFT correlations are doing an excellent job in describing a quasiparticle renormalization with a weight $Z$  that is uniform in momentum space -- a surprisingly  good approximation for many transition metal-oxides and heavy fermion systems; as well as the  Mott-Hubbard metal-insulator transition that emerges for $Z\rightarrow 0$ \cite{Georges1992a}. Furthermore, all kinds of orders (magnetic, orbital, charge density wave $\ldots$) are quite well captured in three-dimensional systems up to the vicinity of the phase transition. Here, close to the phase transition, the mean-field nature of DMFT surfaces, among others, in form of mean-field critical exponents.
Non-local correlations  are here essential for describing the proper critical behavior. With a diverging correlation length,   long-range correlations feed back to the self-energy which thus becomes non-local.

The DMFT approach has been covered already in various other chapters of this Autumn School \cite{Pavarini2022}, and  the present chapter will thus focus instead on non-local electronic correlations beyond DMFT.  When are such non-local correlations important?

\begin{figure}[tb!]
 \centering
 \includegraphics[width=0.6\textwidth]{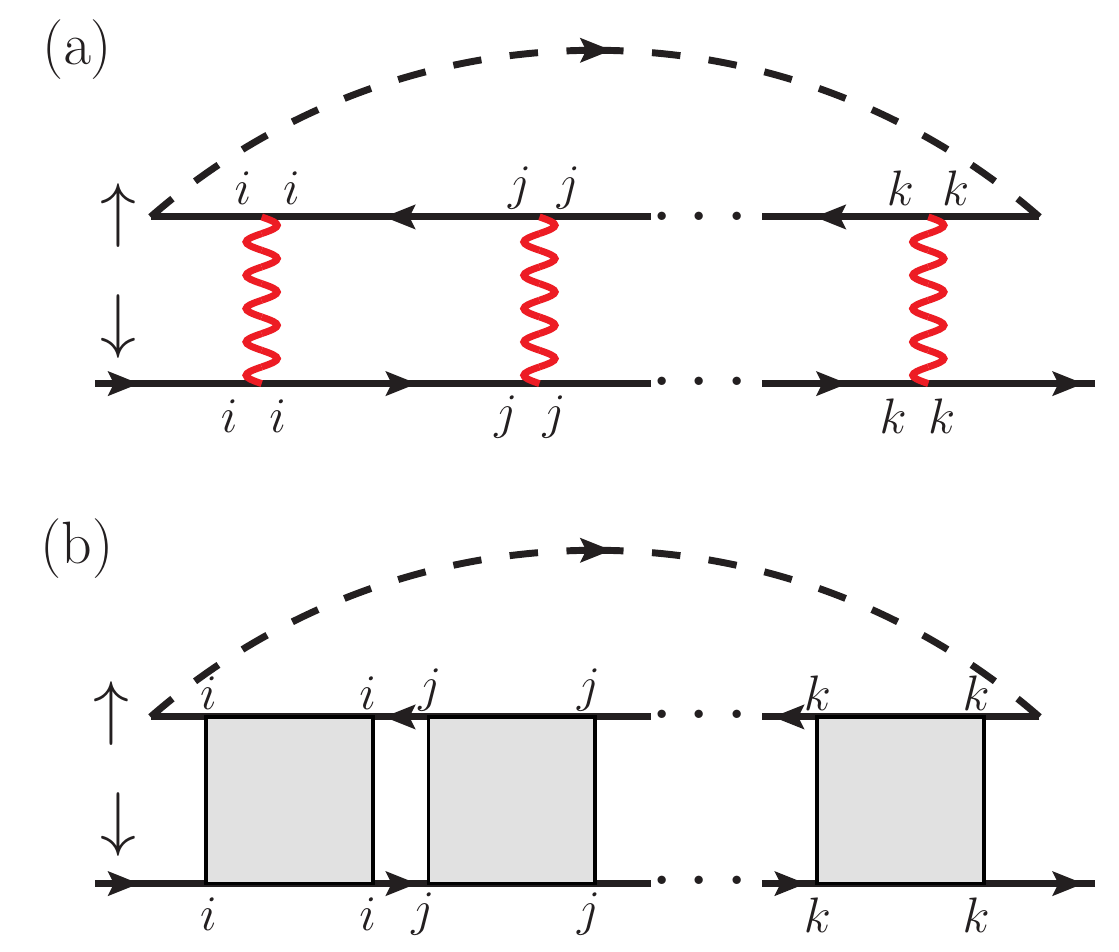}
 \caption{Top: Feedback of spin fluctuations (paramagnons) \index{spin fluctuations} calculated in RPA to the self-energy. Bottom: In diagrammatic extensions of DMFT such physics is taken into account but with the bare interaction $U$ (red wiggled line) replaced by a local vertex \index{vertex} calculated from an impurity model (light gray box). Depending on the flavor of the diagrammatic extension of DMFT different local vertices are employed. Reproduced from \cite{RMPVertex}.}
 \label{Fig:selfenergy}
\end{figure}

A lot of our insight into physical phenomena stems from weak coupling perturbation theory. Even though such an approach is certainly not applicable to strongly correlated electron systems, it often nonetheless provides for some qualitative understanding. One example, where non-local correlations enter the self-energy are spin-fluctuations. These can be calculated at weak coupling in the random phase approximation (RPA) \index{random phase approximation} which is discussed in quantum field theory textbooks.
The RPA is the geometric series of all ladder diagrams with no, one, two etc. Coulomb interaction lines, as illustrated in Fig.~\ref{Fig:selfenergy} (top). The figure just shows one term of the sum, where ``$\ldots$'' indicates that all orders
in $U$ are included.
The RPA can be used to calculate the magnetic susceptibility Fig.~\ref{Fig:selfenergy} (top, without dashed line). As we will see later, the poles of this susceptibility constitute bosonic quasiparticle excitations coined magnons or, in the paramagnetic phase, paramagnons.

Now, when we add the dashed black line in  Fig.~\ref{Fig:selfenergy} (top), we obtain a self-energy Feynman diagram. It describes the coupling of the electron to the spin-fluctuations. We can also call it the electron-(para)magnon interaction. As spin-fluctuations are very non-local we get -- whenever spin-fluctuations become important-- contributions from diagrams with lattice sites  $i\neq j\neq k$ in   Fig.~\ref{Fig:selfenergy} (top). Such self-energy diagrams are certainly not contained in DMFT, which sums up all {\em local} contributions of Feynman diagrams for the self-energy. Hence, whenever spin fluctuations become large, we must expect important corrections to the DMFT self-energy. Indeed this is not restricted to spin fluctuations, but depending on what kind of spin combination the particle and the hole in the RPA ladder have, one can also obtain the coupling of electrons to charge, orbital etc. fluctuations.

That means  non-local correlations are certainly relevant whenever spin, charge etc.\ fluctuations are important. An obvious regime where this is the case is the vicinity of a second-order phase transition as already mentioned.  Here the magnetic, charge etc.\ susceptibility diverges and significant changes to the DMFT solution are thus to be expected. For low dimensional systems we will  get corrections also further away from the phase transition. The Mermin-Wagner theorem prohibits long-range order with a continuous symmetry breaking in two-dimensions at finite temperature. Hence, antiferromagnetic order is restricted to zero temperature. However, above this zero-temperature antiferromagnetic phase, we have now strong antiferromagnetic fluctuations in a wide temperature range, even with exponentially large correlation lengths.
DMFT has been developed with the limit of dimension $d=\infty$ in mind, see \cite{Metzner1989}  and Chapter ``Why calculate in infinite dimensions?'' by D.~Vollhardt \cite{Pavarini2022}. Hence, also from this perspective it is not surprising that we need to expect larger corrections to DMFT for low dimensional systems.

\begin{figure}[t!]
\begin{center} 
\includegraphics[clip=true,width=0.7\textwidth]{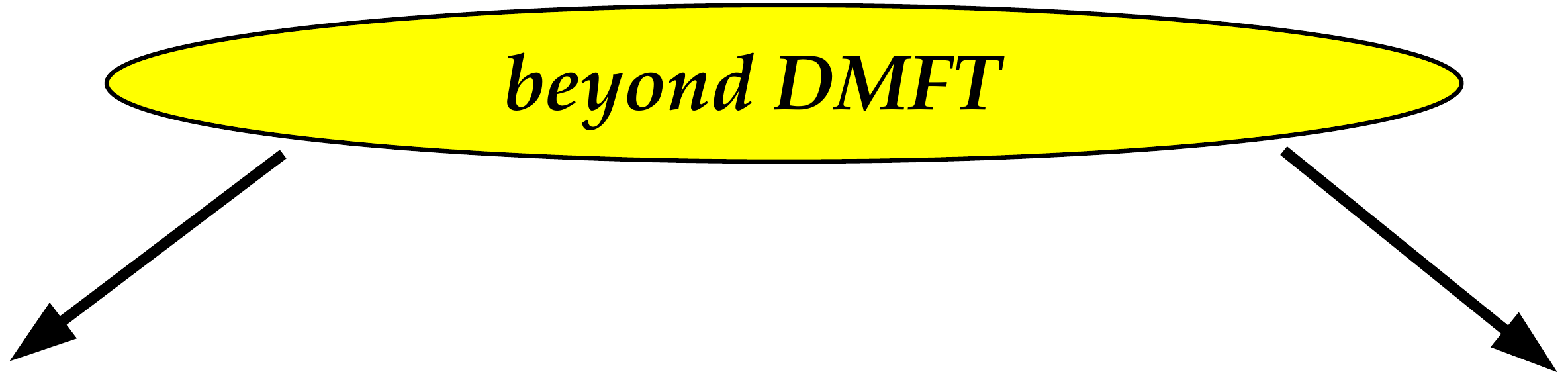}\hspace{1.cm}\phantom{a}
\end{center}
{\bf \large cluster extensions}
\hfill
{\bf \large diagrammatic extensions}
\vspace{.2cm}

\includegraphics[clip=true,width=0.4\textwidth]{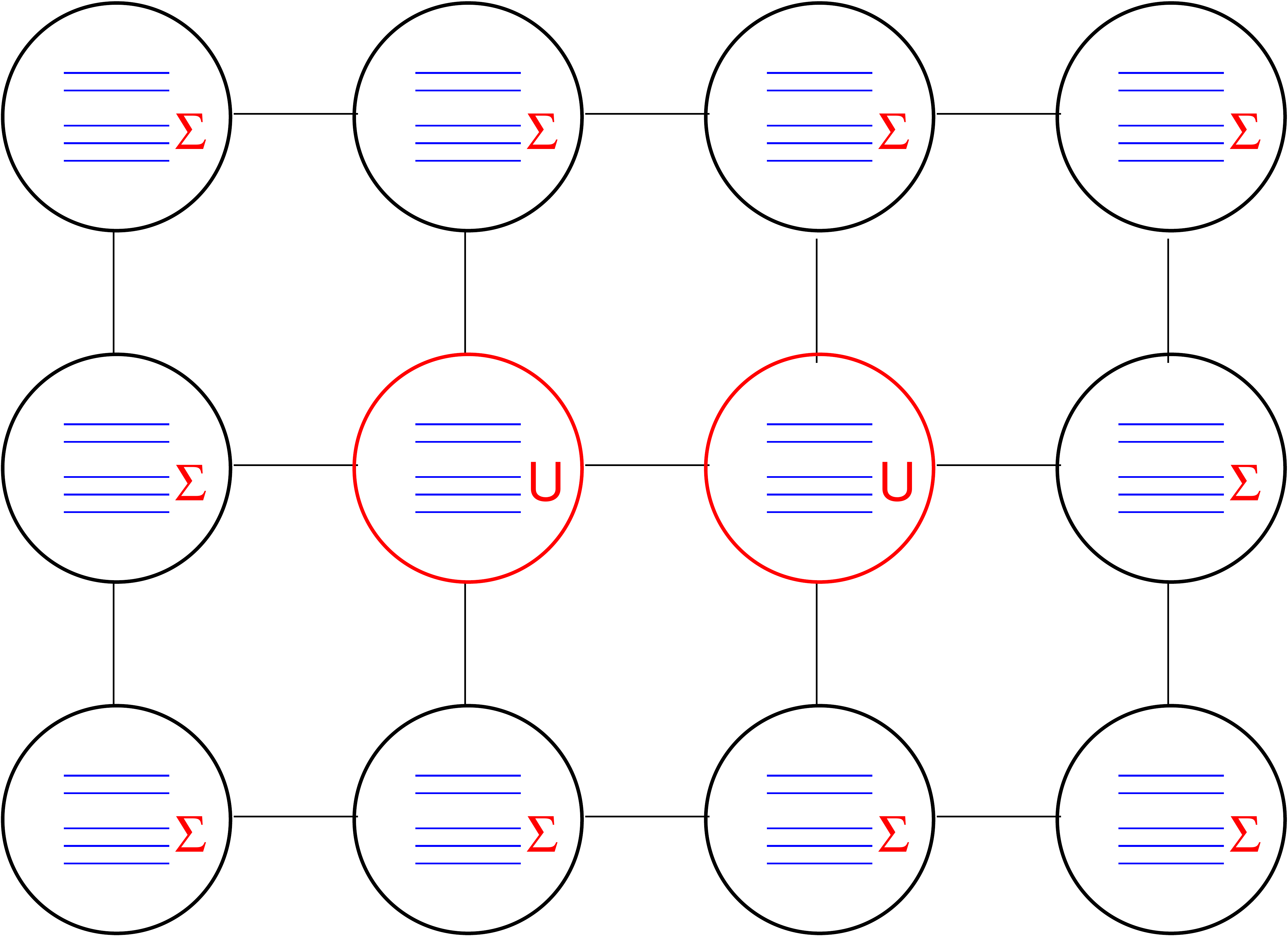}
\hfill 
\includegraphics[clip=false,width=0.5\textwidth]{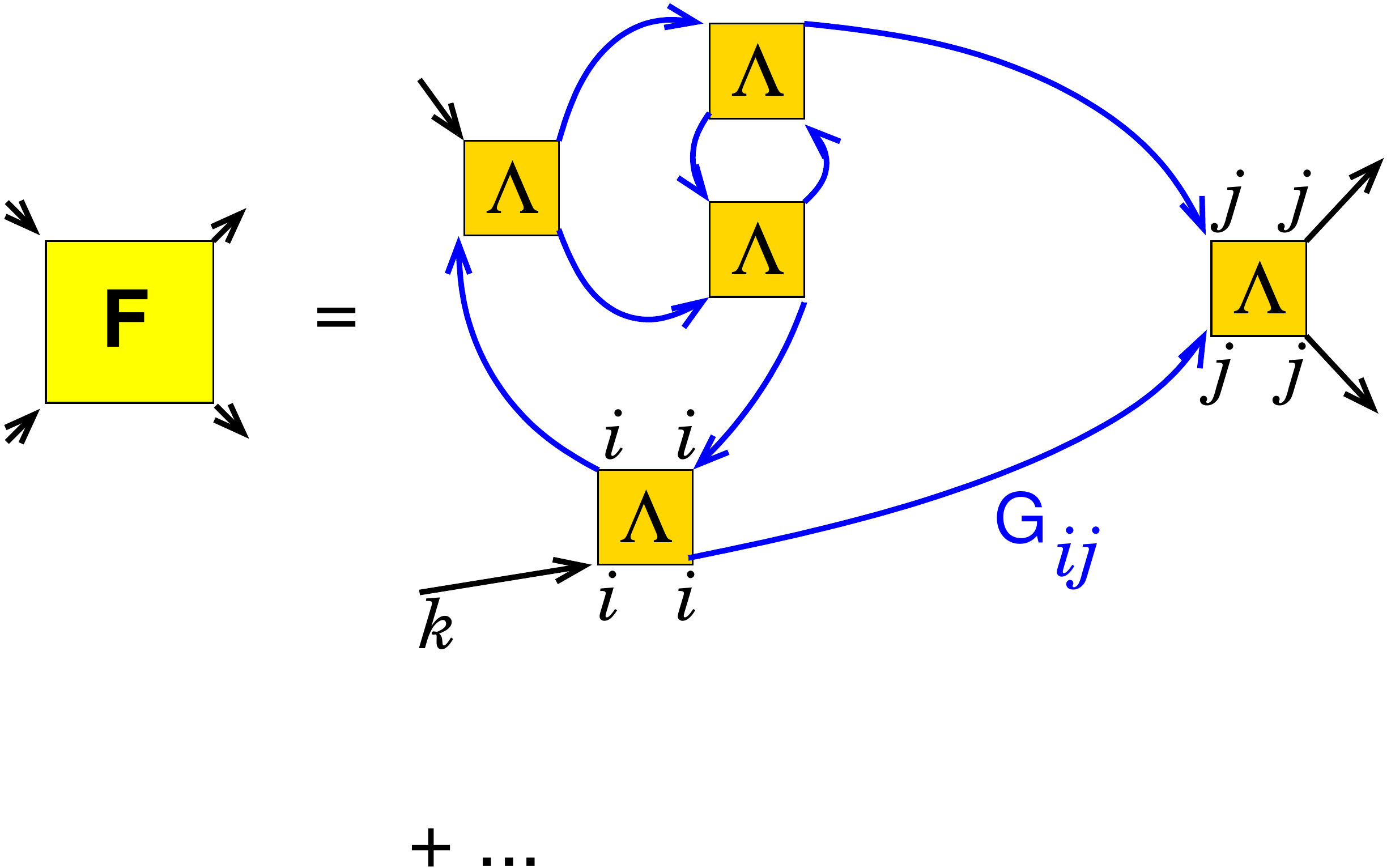}
\vspace{.3cm}

\caption{
  \index{cluster dynamical mean field theory}
  \index{dynamical vertex approximation}
Left: In cluster extensions, a couple of sites  (here two marked in red)
are embedded in a dynamical mean field which can be formulated through
the  self energy $\Sigma$. Restricted to a single site,  DMFT is obtained; extended to the full lattice the exact solution is recovered.
Right: In diagrammatic extensions, the locality of DMFT is put to the next level, the two-particle level. Local two-particle building blocks $\Lambda$ are connected by non-local Green function lines $G_{ij}$, resulting in a non-local full vertex $F$ or susceptibility $\chi$. When restricting $G_{ij}$ to its local contribution
$G_{ii}$,  DMFT is recovered. Adapted from \cite{Held2008}.
\label{Fig:2ways}}
\end{figure}

On the other hand, we would like to keep the success of DMFT in describing a major part of electronic correlations rather well: the local correlations.
Two major routes to do so have been developed to this end, see Fig.~\ref{Fig:2ways} for an illustration.  Cluster extensions of DMFT \cite{Maier2005}
\index{cluster dynamical mean field theory}
put the DMFT concept of locality onto a cluster (instead of a single site) that is embedded in a DMFT bath. For a single site cluster this is just DMFT. Illustrated in  Fig.~\ref{Fig:2ways} is a two-site cluster where thus non-local correlations between the two red sites of the cluster are captured. Such two site clusters can e.g.~describe the formation of a spin singlet between  two electrons on the two sites. For a four site cluster also $d$-wave superconductivity can be described. However such small clusters tend, in practice, to largely overestimate the physics that is compatible with the cluster such as the spin singlet formation and $d$-wave superconductivity for a two and four site cluster, respectively. While one can go to clusters of size $10 \times 10$, a proper finite size scaling remains  challenging. This is even more true for realistic multi-orbital calculations that are restricted to a handful of sites.

The other route extends DMFT \cite{RMPVertex} Feynman diagrammatically. Here, the concept of locality is not extended   to a cluster but instead to the $n$-particle vertex. For $n=1$ we have the one-particle vertex which is nothing but the self-energy. And a local self-energy is just the DMFT approximation.  For $n=2$, i.e., the two-particle level, we start instead with a local two-particle  vertex and construct from it the non-local full vertex $F$, see  Fig.~\ref{Fig:2ways} (right), as well as the non-local self-energy and Green's function. This is the level most commonly applied nowadays. Similarly as for the cluster extensions there are different flavors depending on which local vertex  and which connecting Green's functions are taken. At the end of the day, most of these different flavors are very similar, at least as long as they take the two-particle, four-point vertex as a starting point.
Other approaches start from a three-point local vertex which is a  more severe approximation.
For an overview and comparison, see the review  \cite{RMPVertex}.

The originary method, called dynamical vertex approximation (D$\mathrm \Gamma$A) \cite{Toschi2007} \index{dynamical vertex approximation}, considers the vertex in terms of real fermions as local. A second widely applied approach is the dual fermion (DF) approach \cite{Rubtsov2008} \index{dual fermions}.
Also within D$\mathrm \Gamma$A different flavors are used.
In its most complete form,  the fully irreducible vertex $\Lambda$ is approximated to be local. Then the parquet equations are needed to construct $F$ and the self-energy $\Sigma$.
 A local  $\Lambda$ is an excellent approximation, even for the two-dimensional Hubbard model in the superconducting doping regime \cite{Maier2006}.
 In the  ladder D$\mathrm \Gamma$A variant, the irreducible vertex $\Gamma_\ell$ in a certain channel $\ell$ such as the particle-hole ($ph$) channel is considered to be local. In this case, the Bethe-Salpeter equation is sufficient to calculate $F$ along the same lines as in the RPA, only now with $\Gamma_\ell$ instead of $U$ as a building block, i.e the light gray box in  Fig.~\ref{Fig:selfenergy} (bottom) is $\Gamma_\ell$ in this case. The ladder variant is  numerically much less expensive. It is sufficient if a certain channel dominates, but does not capture the coupling of different channels into each other.

 One can also extend the concept of locality to higher $n$-particle vertices. The $n\!=\!3$-particle vertex level, for example, has been employed for estimating the error of the $n\!=\!2$-particle calculations \cite{Ribic2017b}. For $n\rightarrow \infty$ the exact solution is recovered.

 Turning back to  Fig.~\ref{Fig:selfenergy}, we see that such diagrammatic extensions are well suited to describe spin-fluctuations and their feedback to the fermionic self-energy.
 The same physics as is qualitatively described in RPA, is now captured for strongly correlated electrons since the local $\Lambda$ already
encodes non-perturbatively all DMFT correlations. 
 The Bethe-Salpeter ladder  of Fig.~\ref{Fig:selfenergy} is precisely the same as is also used to calculate the DMFT susceptibility, cf.~the Chapter ``DMFT for response functions'' by E.~Pavarini   \cite{Pavarini2022}. What is not covered in DMFT is how these spin-fluctuations impact the self-energy as well as self-consistency effects, i.e., how the changed $\Sigma$ modifies $G$   or that the local $\Gamma_\ell$ itself becomes different from DMFT. These self-consistency effects lead to a dampening of the spin-fluctuations compared to the mean-field DMFT solution, up to the point of fulfilling the Mermin-Wagner theorem in two-dimensions.

Diagrammatic extensions of DMFT have been highly successful: (i) The critical behavior in the vicinity of  (quantum) phase transition  could be described for the first time in electronic models, a topic which was covered already excessively in the Autumn School 2018 \cite{Held2018}, see also the review~\cite{RMPVertex}.
(ii) It was realized that the two-dimensional square lattice Hubbard model with perfect nesting  is insulating all the way down to zero interaction  \cite{Schaefer2015-2}, correcting earlier cluster DMFT results. (iii) The pseudogap and $d$-wave superconductivity can be described in the two-dimensional Hubbard model, a topic which we will cover in the following chapters. (iv) A new polariton, the $\pi$-ton has been discovered in model calculations \cite{Kauch2020}. (v) Realistic materials calculations are possible and have been pursued e.g.\ for SrVO$_3$ \cite{Galler2016}.

In the following, we will first  introduce one the diagrammatic extensions of DMFT,  the D$\mathrm \Gamma$A, in Sec.~\ref{Sec:DGA}. Simplifications when using the  ladder variant of D$\mathrm \Gamma$A are outlined in Sec.~\ref{Sec:ladderDGA}. The Hubbard model is introduced in Sec.~\ref{Sec:model} and its justification for cuprates and nickelates is discussed.  Physical results regarding spin fluctuations, the pseudogap and superconductivity are discussed in Sec.~\ref{Sec:spin fluct}, Sec.~\ref{Sec:PG}, and Sec.~\ref{Sec:SC}, respectively.
Finally, Sec.~\ref{Sec:conclusion} provides a conclusion and outlook.

\section{Dynamical vertex approximation}
\label{Sec:DGA}
 \index{dynamical vertex approximation}
 The aim of the present section is to provide the reader with the basic  idea of the dynamical vertex approximation. It builds upon a similar chapter of a preceding J\"ulich Autumn School \cite{Held2018}. A more detailed description for a second reading can be found in the review  \cite{RMPVertex}.
Further information on how to calculate the superconducting critical temperature $T_c$ and how to include the asymptotic form of the vertex, can be found in   \cite{Kitatani2022}.

The basic idea of the dynamical vertex approximation (D${\rm\Gamma}$A) is a resummation of Feynman diagrams, not order by order of the Coulomb interaction as in conventional perturbation theory, but in terms of their locality. That is, we assume the fully irreducible $n$-particle vertex to be local and from this building block we construct further diagrams and non-local correlations.

\begin{figure}[t!]
 \centering \includegraphics[width=\textwidth]{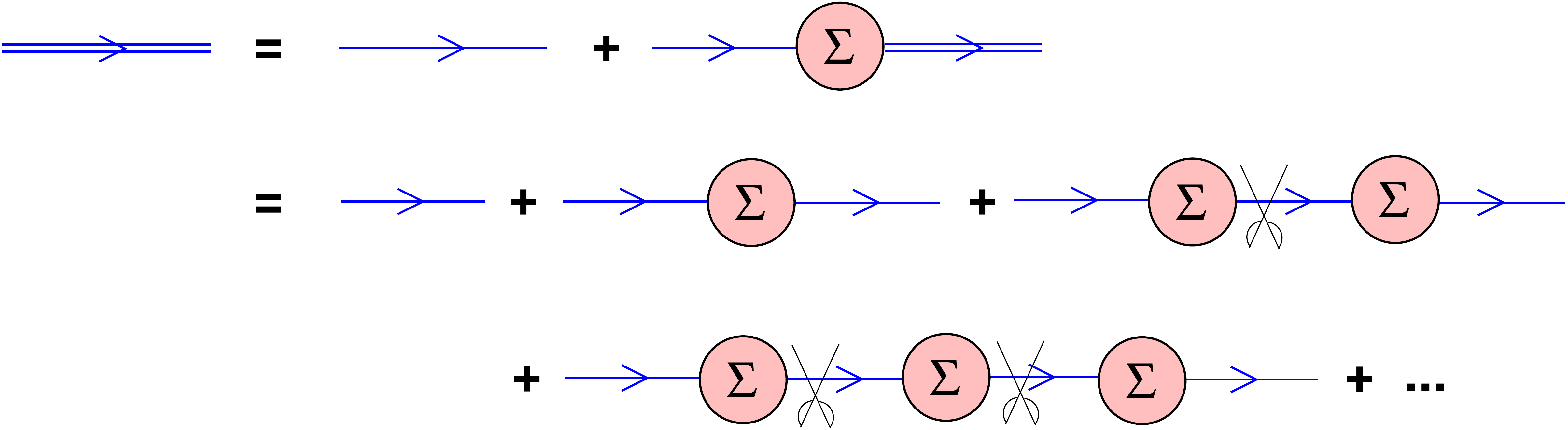}
 \caption{Dyson equation connecting the Green function and self energy $\Sigma$ (single blue line: non-interacting  $G_0$; double blue line: interacting $G$). 
The pair of scissors indicates that these diagrams are one-particle reducible (i.e., cutting one $G_0$ line separates the Feynman diagram into two parts). From \cite{Held2018}. \label{Fig:DGAF2}
\index{Dyson equation}}
\end{figure}

The first level ($n=1$)  is then just the DMFT which  corresponds to all local Feynman diagrams for the self-energy $\Sigma$. Note that $\Sigma$ is nothing but the fully irreducible $n\!=\!1$-particle vertex. One particle-irreducibility here means that cutting one Green's function line does not separate the Feynman diagram into two pieces. Indeed such reducible diagrams must not be included in the self-energy since it is exactly these diagrams that are generated from the Dyson equation (Fig.~\ref{Fig:DGAF2}) which resolved for $G$ reads
 \begin{eqnarray}
G_{{\mathbf k} \nu}=[1/G^0_{{\mathbf k}\nu}-\Sigma_{{\mathbf k}\nu}]^{-1} 
\label{Eq:Dyson}
\index{Dyson equation}
\end{eqnarray}
for momentum ${\mathbf k}$, Matsubara frequency $\nu$ and non-interacting Green's function $G^0_{{\mathbf k} \nu}$. Here and in the following, single blue lines denote non-interacting Green's functions $G_0$ and double blue lines indicate interacting Green's functions $G$.  Fig.~\ref{Fig:DGAF2}  further  shows  how  one-particle reducible diagrams are generated through the Dyson equation. Hence these must not be contained in the Feynman diagrams that constitute $\Sigma$, to avoid a double counting. That means, $\Sigma$ must be one-particle reducible in terms of $G^0$.\footnote{In terms of $G$ the skeleton diagrams for $\Sigma$ are also two-particle reducible. But that is another story that is connected with the way how $\Sigma$ enters $G$.}

\begin{figure}[t!]
  \includegraphics[width=\textwidth]{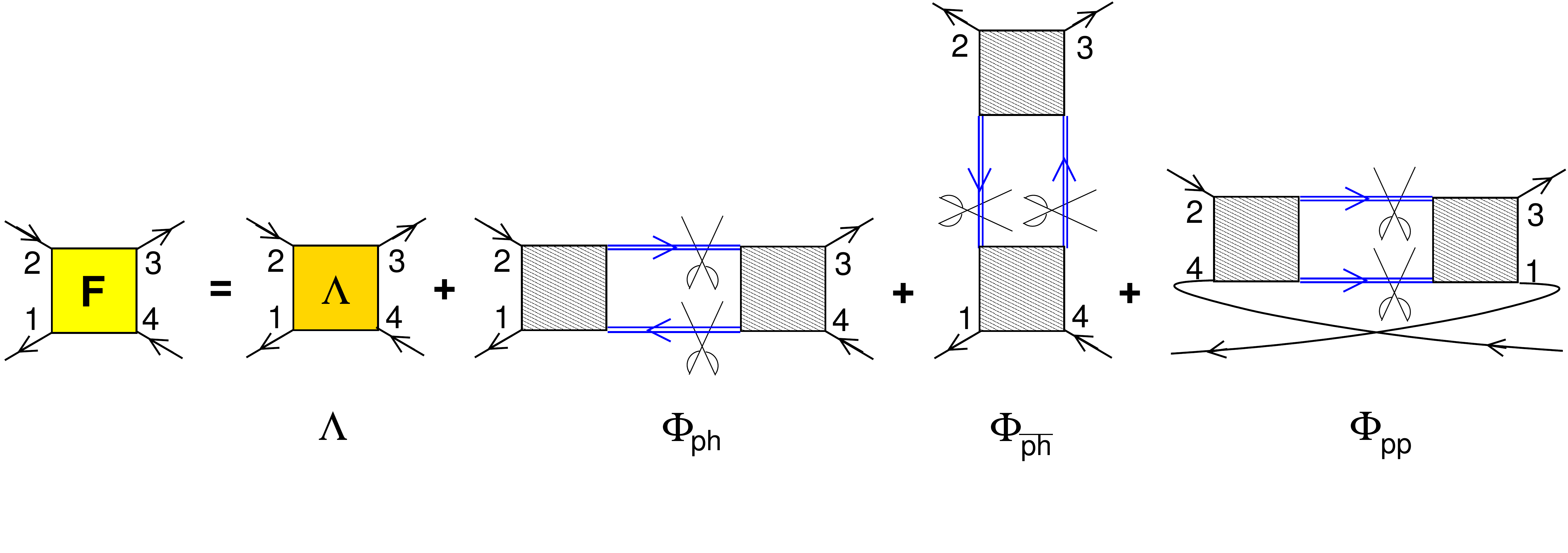}
  \vspace{-.5cm}
 \caption{Parquet decomposition of the full (reducible) vertex $F$ into 
 the fully irreducible
 vertex $\Lambda$ and two-particle reducible diagrams $\Phi_\ell$ in the three channels. The two pairs of scissors indicate the reducibility of the three $\Phi_\ell$'s. Each two-fermion Feynman diagram belongs to one and only one of the three  $\Phi_\ell$'s or to $\Lambda$. \label{Fig:DGAF5}
\index{parquet equation} }
\end{figure}

On the next level, for $n=2$, we assume the locality of the two-particle  fully irreducible vertex $\Lambda$. For the two-particle vertex, fully irreducible means that  cutting two Green's function lines does not separate the diagram into two pieces.
There are three different kinds (channels) $\ell$ of reducible vertices  $\Phi_\ell$ and a fourth,  $\Lambda$, that is fully irreducible. Most importantly each diagram falls in one and only one of those four subgroups. Thus the full vertex $F$, containing all diagrams, can be written as the sum those. This decomposition of the vertex is called \textit{parquet decomposition} and is graphically displayed in Fig.\,\ref{Fig:DGAF5}.

The reason why there  are 
three distinct reducible parts $\Phi_\ell$ is that say leg $1$ may stay connected with leg $2$, $3$, or $4$ when cutting two Green's function lines as indicated in  Fig.~\ref{Fig:DGAF5}. 
These three possible   channels $\ell$ are denoted as   particle-hole ($ph$), transversal particle-hole ($\overline{ph}$) and particle-particle ($pp$).
The irreducible vertex of each channel is just the complement: 
$\Gamma_\ell=F-\Phi_\ell$.
It is important to note that each reducible diagram is contained in one and only one of these channels. One can show this by contradiction: otherwise cutting lines in two channels  would result in a diagram with one incoming and two outgoing lines, which is not possible because of the conservation of (fermionic) particles.

There is a set of six exact equations, also called the ``parquet equations'' \cite{Bickers2004,RMPVertex} to the confusion  of the common  student, that allows us to calculate from a given $\Lambda$ the six quantities: full vertex $F$,  self-energy $\Sigma$, Green's function $G$ and the three reducible vertices $\Phi_\ell$.  

{\em (1)} The first equation is the actual parquet equation  [Fig.~\ref{Fig:DGAF5}, Eq.~(\ref{eq:parquet})]
\begin{eqnarray} 
 F_{r,\mathbf{k}\mathbf{k'}\mathbf{q}}^{\nu\nu'\omega}&=&
 \Lambda_{r,\mathbf{k}\mathbf{k'}\mathbf{q}}^{\nu\nu'\omega}+
 \Phi_{ph,r,\mathbf{k}\mathbf{k'}\mathbf{q}}^{\nu\nu'\omega}+ 
 \Phi_{\overline{ph},r,\mathbf{k}\mathbf{k'}\mathbf{q}}^{\nu\nu'\omega}+
 \Phi_{pp,r,\mathbf{k}\mathbf{k'}\mathbf{q}}^{\nu\nu'\omega}
 \label{eq:parquet}
\index{parquet equation}
\end{eqnarray}
where  $r\in\{c,s\}$ is the  symmetric/antisymmetric  spin combination, i.e., $F_{c/s}= F_{\uparrow  \uparrow}
  \pm  F_{\uparrow  \downarrow} $ and similarly for other vertices.\footnote{This assumes SU(2) symmetry, if this is broken altogether four spin combinations need to be taken into account.} The name indicates that $F_{c}$ and $F_{s}$ give rise to the charge and spin fluctuations, respectively.
  
\begin{figure}[t!]
  \begin{center}
    \includegraphics[width=.8\textwidth]{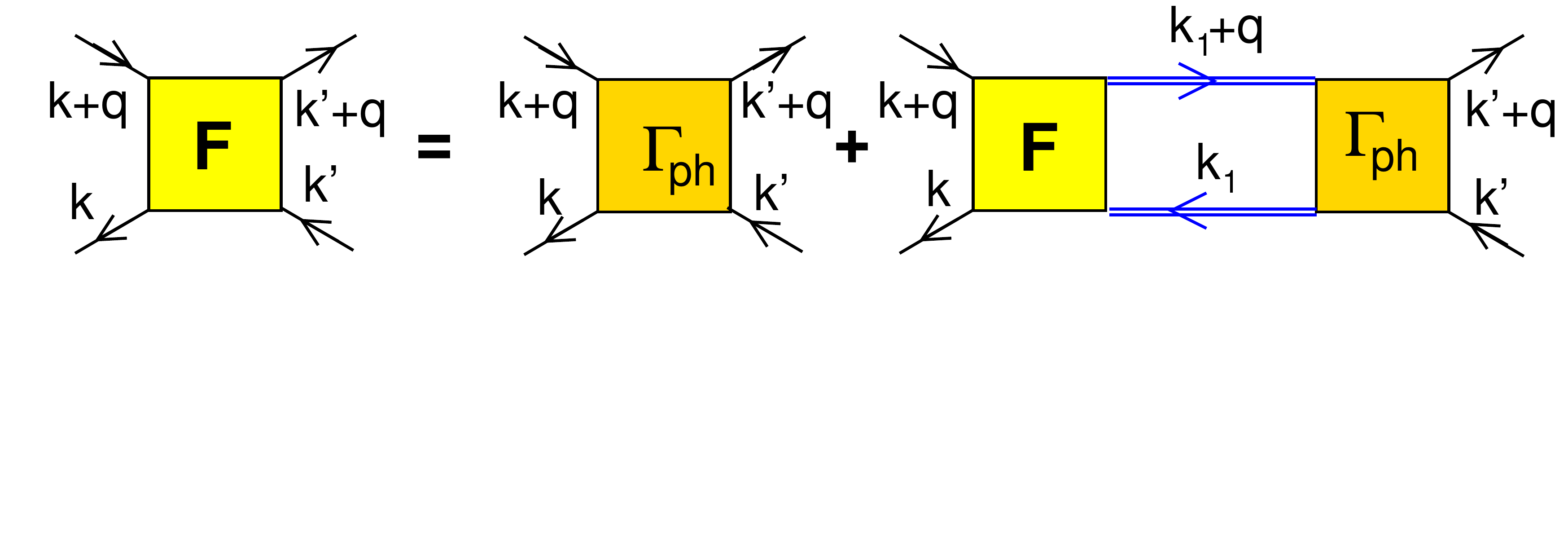}
  \end{center}
  \vspace{-2.4cm}
  
  \caption{Bethe-Salpeter equation in the particle-hole channel, $\ell=ph$, allowing us to calculate $F$ from the irreducible vertex in $\Gamma_{ph}$ in this channel. The rightmost term produces the particle-hole reducible diagrams denoted as  $\Phi_{ph}$ in Fig.~\ref{Fig:DGAF5};  $\Gamma_{ph}=\Lambda+ \Phi_{\overline{ph}} + \Phi_{pp}$.\label{Fig:BSE}\index{Bethe-Salpeter equation} }
\end{figure}

  {\em (2-4)} Three of the six ``parquet'' equations   are the Bethe-Salpeter equation in the three channels $\ell$.
  In [Fig.~\ref{Fig:BSE}, Eq.~(\ref{eq:BSE})] we here only reproduce the $\ell=ph$ channel; again with $r\in\{c,s\}$ for a symmetric/antisymmetric spin combination\footnote{We implicitly assume a proper normalization of the momentum and Matsubara frequency sums, i.e., $\sum_{\mathbf{k}}\widehat{=}\frac{1}{N_K}\sum_{\mathbf{k}}$ and $\sum_{\nu}\widehat{=}\frac{1}{\beta}\sum_\nu$, where $N_k$ is the number of momentum points and $\beta=1/T$ the inverse temperature.}
\begin{align} 
 \label{eq:BSE} 
  F_{r,\mathbf{k}\mathbf{k'}\mathbf{q}}^{\nu\nu'\omega}=\Gamma_{r,ph,\mathbf{k}\mathbf{k'}\mathbf{q}}^{\nu\nu'\omega}+\sum_{\mathbf{k_1}\nu_1}F_{r,\mathbf{k}\mathbf{k_1}\mathbf{q}}^{\nu\nu_1\omega} G_{\mathbf{k_1}\nu_1}G_{(\mathbf{k_1}+\mathbf{q})(\nu_1+\omega)}\Gamma_{ph,r,\mathbf{k_1}\mathbf{k'}\mathbf{q}}^{\nu_1\nu'\omega}\, ;
\index{Bethe-Salpeter equation}
\end{align}
One often combines  frequency $\nu$ and momentum $\mathbf k$ to a four vector $k=({\mathbf k}, \nu)$. We do not do so in this Chapter for the equations, but  in the figures $k$ represents $({\mathbf k}, \nu)$.
Further it is important to choose a momentum-frequency convention for the vertices and stick to it. Because of energy and momentum conservation we only need three momenta and frequencies for the four points $1,2,3,4$ of the two-particle vertices in Fig.~\ref{Fig:DGAF5}. Unless noted otherwise, we use the $ph$ convention which is the natural one for the $ph$ channel and already used in Fig.~\ref{Fig:BSE}.
That is, the $1,2,3,4$ frequency-momenta of Fig.~\ref{Fig:DGAF5} are related to Fig.~\ref{Fig:BSE} as follows: $k_1=k$, $k_4=k^\prime$, $k_2=k+q$, and---because of energy-momentum conservation--- $k_3=k^\prime+q$. 
The Bethe-Salpeter equations in the other channels
have the same structure just with another $\Gamma_\ell$, $\ell=\overline{ph}$ or $pp$, with  another way to connect the building blocks (see Fig.~\ref{Fig:DGAF5}), and with another natural (diagonal) frequency-momentum $q$.

{\em (5)} The fifth equation is the  Dyson equation that we already introduced  [Fig.~\ref{Fig:DGAF2}, Eq.~(\ref{Eq:Dyson})]. 

\begin{figure}[t!]
  \begin{center}
    \includegraphics[width=.5\textwidth]{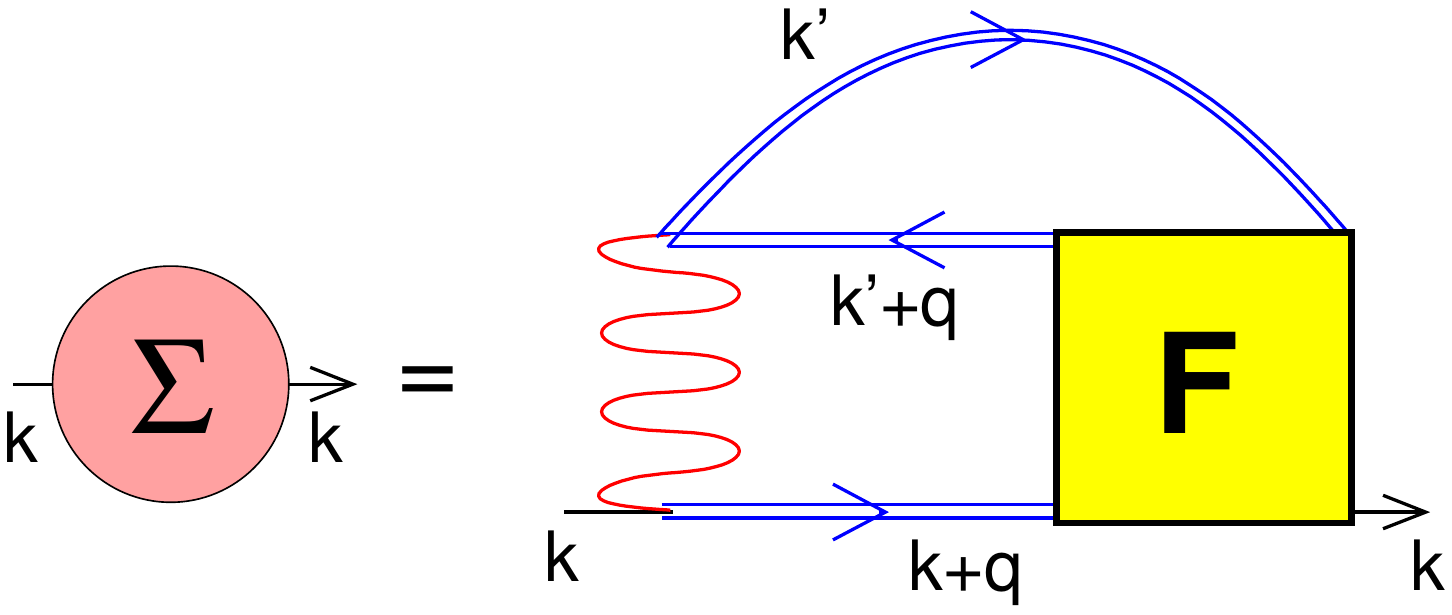}
  \end{center}
  \vspace{-.4cm}
  
 \caption{Schwinger-Dyson equation connecting the full vertex $F$ and the self-energy $\Sigma$.\label{Fig:SDE} \index{Schwinger-Dyson equation}}
\end{figure}

{\em (6)} Finally, the sixth equation is the Schwinger-Dyson  equation [Fig.~\ref{Fig:SDE}, Eq.~(\ref{eq:SD})] which reads
\begin{align}\label{eq:SD}
  \Sigma_{ {\mathbf k} \nu} &= \frac{U n}{2}-\frac{U}{2}\sum_{k^{\prime},q}\sum_{\nu^{\prime},\omega}\left[F_{{c}, {\mathbf k \mathbf k^{\prime}\mathbf q}}^{\nu \nu^{\prime}\omega}-F_{s, {\mathbf k \mathbf k^{\prime}\mathbf q}}^{\nu \nu^{\prime}\omega}\right] G_{({\mathbf k+ \mathbf q})(\nu+\omega)}G_{{\mathbf  k^{\prime}}\nu^{\prime}}G_{({\mathbf k^{\prime}+\mathbf q})(\nu^{\prime}+\omega) }\;
  \index{Schwinger-Dyson equation}
\end{align}
and  connects $\Sigma$ and $F$. Here we consider a single-orbital and local interaction $U$ that connects opposite spins. For this reason only $F_{\uparrow\downarrow}=(F_c-F_s)/2$ enters. The first term is simply the Hartree(-Fock) term, which is not included in Fig.~\ref{Fig:SDE}  and where $n$ is the average number of electrons per site in the paramagnetic phase.

\begin{figure}[t!]
  \centering \includegraphics[width=0.9\textwidth]{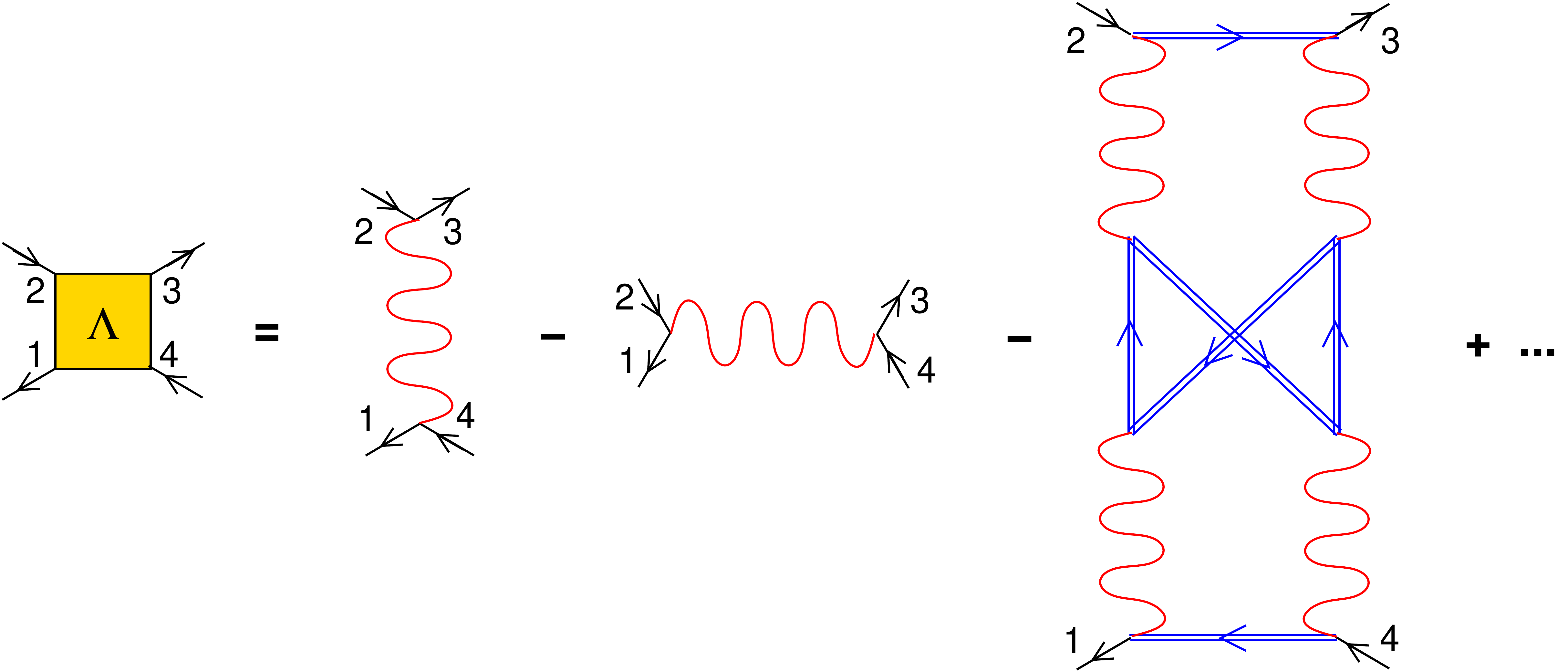}
 \caption{Lowest order Feynman diagrams that contribute to the fully irreducible vertex $\Lambda$. \index{vertex}
 \label{Fig:lambda}}
\end{figure}

\index{parquet equation}
This set of six exact parquet equations allows us to determine the six quantities $F$, $\Phi_\ell$, $\Sigma$, and $G$, as well as 
$\Gamma_\ell=F-\Phi_\ell$. If we knew the exact $\Lambda$, we could determine all of these quantities exactly, as well as associated one- and two-particle physics. Unfortunately, we do not know the exact $\Lambda$. Hence we need some approximation.
If we approximate $\Lambda$
by the bare Coulomb interaction $U$, we obtain the parquet approximation \cite{Bickers2004}. We can do better than this, and in D$\mathrm \Gamma$A we approximate
$\Lambda$ by all local Feynman diagrams.
Quantum Monte Carlo simulations show that this is an excellent approximation \cite{Maier2006}. Indeed $\Lambda$ is very compact. All two-particle reducible diagrams are generated from it and the first irreducible diagram that enters $\Lambda$ besides the bare Coulomb interaction is of fourth order, see Fig.~\ref{Fig:lambda}. So to speak the irreducible diagrams  $\Lambda$ form a skeleton from which many more diagrams are generated. Because of this, it is more local than $\Gamma_\ell$ and, in particular, much more local than $F$, $\Sigma$ or $G$.

This local irreducible $\Lambda$ is ``summed up'' in practice by solving an Anderson impurity model, similar as in DMFT but now we calculate the two-particle Green's function of the impurity model  and from this $\Lambda$.
For example, we can use continuous-time quantum Monte Carlo simulations in the hybridization expansion see "QMC as a solver for DMFT'' by P.~Werner  \cite{Pavarini2022}, e.g.\ using the {\tt w2dynamics} package \cite{w2dynamics} which allows us to calculate all two-particle responses using worm sampling \cite{Gunacker15}.

In principle, one can then further turn to the $n\!=\!3$-particle level etc.; and, for the  $n\rightarrow \infty$-particle  fully irreducible local vertex D${\rm\Gamma}$A one recovers the exact solution.
As a matter of course determining the $n\!=\!3$-particle vertex becomes already cumbersome. But it may serve
at least as an  error estimate \cite{Ribic2017b} if one is truncating the scheme at the two-particle vertex level. Also completely new physics, that is hitherto not understood, may be hiding behind diagrams generated from the $n\!=\!3$-particle irreducible vertex. Some physical processes such as Raman scattering naturally require three-particle vertex functions.

In the the present section we have learned about the fully fledged parquet  D$\mathrm \Gamma$A. It is unbiased with respect to all three channels $\ell$ and thus treats antiferromagnetic or charge fluctuations in the $ph$-channel on a par with
e.g.\ superconducting fluctuations or weak localization corrections in the $pp$ channel.
It also mixes the different channels. For two channels this mixing generates, with some fantasy, a traditional parquet floor like pattern with ladder rungs in one direction intermixed by ladder rungs in the orthogonal direction etc., hence the name.
Being unbiased definitely is a great advantage. For example,
in \cite{Kauch2020} we were looking for weak localization corrections to the optical conductivity in the $pp$ channel and excitons that live in the $ph$ channel. Instead we found that for strongly correlated systems that host strong alternating spin or charge fluctuations, the $\overline{ph}$ channel is actually dominant for the optical conductivity, against all expectations.

The drawback of the full parquet solution is that we deal with three momenta and three frequencies. Even on a rather coarse discretization grid, we thus easily end up with Terabytes of data. In the  Bethe-Salpeter equation (\ref{eq:BSE}) the  bosonic momentum $\mathbf {q}$ and frequency $\omega$ decouple, so that it can be well distributed on several computer cores without the need to communicate.
However, this natural  $\mathbf {q}$ and $\omega$ is channel dependent (see \cite{RMPVertex}), i.e., different for  the three channels of Fig.~\ref{Fig:DGAF5}.
Hence, when we add the three channels in the  parquet equation (\ref{eq:parquet}), we mix different momenta, requiring a lot of communication between different cores, which --say-- have previously solved the Bathe-Salpeter equation in different channels. Network traffic thus becomes the computational bottleneck on a supercomputer for parquet D$\mathrm \Gamma$A. Efforts to mitigate this problem are the truncated unity basis
for momentum space \cite{Eckhardt2020}, the intermediate representation (IR) 
for frequency space \cite{Wallerberger2021}, and the single boson exchange decomposition \cite{Krien2019}.

\section{Ladder dynamical vertex approximation}
\label{Sec:ladderDGA}
\index{dynamical vertex approximation}

\begin{figure}[t!]
  \begin{center}
    \includegraphics[width=.85\textwidth]{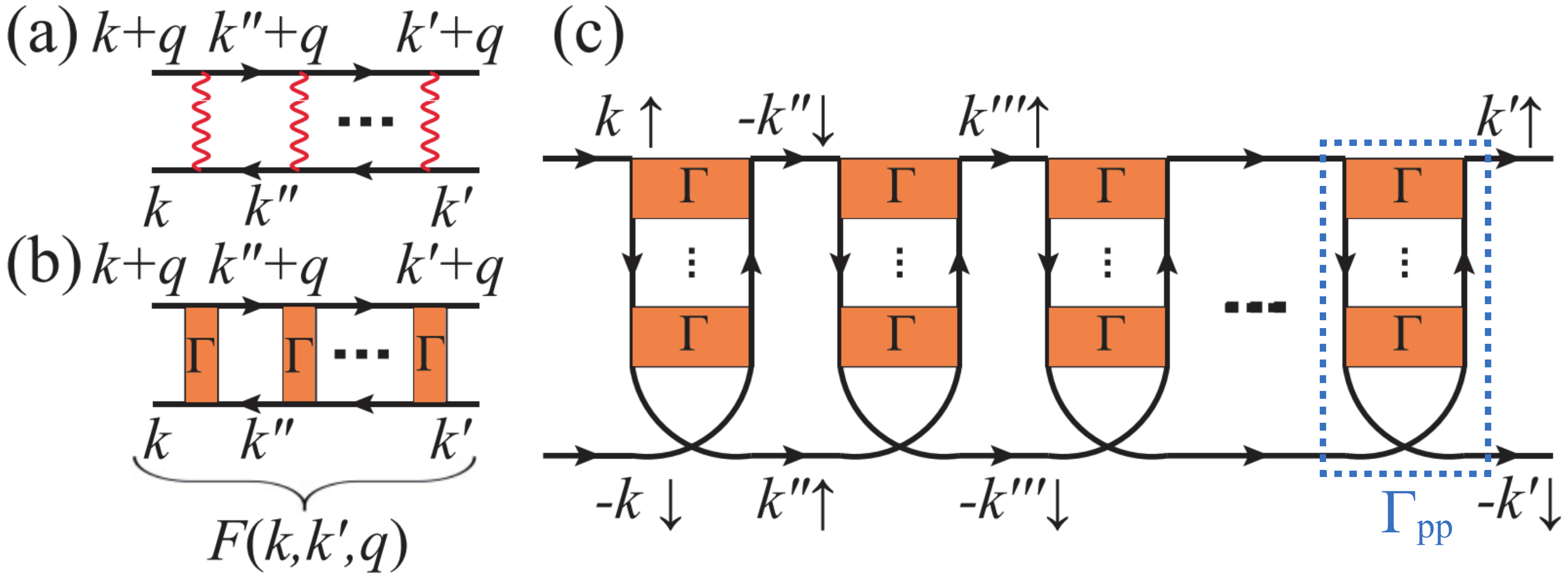}
  \end{center}
\caption{(a) Spin fluctuations as calculated from the Bethe-Salpeter ladder diagrams in terms of the bare Coulomb interaction $U$ in RPA. (b) Same as (a) but now calculated in ladder D$\mathrm \Gamma$A from the local irreducible vertex in the $ph$ channel: $\Gamma\equiv \Gamma_{ph}$. (c) These antiferromagnetic spin fluctuations enter 
as $\Gamma_{pp}$ (via the parquet equation) in the particle-particle channel. This leads to the binding of two electrons into a Cooper pair,
and if this bosonic Cooper pair Bose-Einstein condensates to superconductivity.
From \cite{Kitatani2019}.\label{Fig:ladderDGA}}
\end{figure}

If we know that a certain channel is dominating, we can restrict ourselves to this particular channel and neglect the others and the mixing of different channels through the parquet equation.
Since  $\Phi_{ph}$ and $\Phi_{\overline{ph}}$  are related by crossing symmetry (invariance of exchanging the two incoming lines \cite{RMPVertex}) both of these channels contribute to $F$ in the same way, albeit with  a crossing-symmetry exchanged frequency and momentum combination. 
Hence, both channels must be included in general.\footnote{When we eventually connect $F$ with altogether four Green's functions to a susceptibility
with a single bosonic momentum and frequency [see Eq.~(\ref{Eq:sus}) and Fig.~\ref{Fig:sus} below)], the ${\mathbf q}=(\pi,\pi,\ldots)$ susceptibility is dominated
 in case of antiferromagnetic spin fluctuations by the $ph$ channel, whereas the  ${\mathbf q}=(0,0,\ldots)$ optical susceptibility (conductivity) is dominated
by the transversal particle-hole channel ($\pi$-tons\cite{Kauch2020}).}
In this case, we hence only need to solve the  Bethe-Salpeter equation (\ref{eq:BSE})  and can consider a local $\Gamma_{ph}$ as input. For a bare interaction $U$, the  Bethe-Salpeter equation (\ref{eq:BSE})  yields diagrams as in Fig.~\ref{Fig:ladderDGA} (a). If we improve on this and use a    local $\Gamma_{ph}$ we get the ladder D$\mathrm \Gamma$A approach and the diagrams  of Fig.~\ref{Fig:ladderDGA} (b).
From these diagrams (plus the contribution from the crossing-symmetrically related  $\overline{ph}$ channel) we get $F$, and from $F$ in turn through the Schwinger-Dyson equation (\ref{eq:SD}) the self-energy.

If we want to calculate superconducting fluctuations in  ladder D$\mathrm \Gamma$A we can plug the antiferromagnetic spin fluctuations as a superconducting pairing glue (which is nothing but $\Gamma_{pp}$) into the $pp$ channel. This is so-to-speak a poor man's one-step parquet calculation. We get the $ph$ and $\overline{ph}$  spin fluctuations into the $pp$ channel, but we do not feed back the $pp$ fluctuations to the
$ph$ and  $\overline{ph}$ channel.  For details on the superconducting calculations and also regarding the treatment of high frequencies, see   \cite{Kitatani2022}.

A self-consistency with respect to the recalculation of the Green's function entering the   Bethe-Salpeter equation (\ref{eq:BSE}) is possible \cite{Kaufmann2021}. Different schemes have been proposed  as well for a self-consistency  with respect to $\Gamma$ (or $\Lambda$) starting with \cite{Quadrilex}, for an overview see \cite{RMPVertex}. A simpler and widely used approach is to do, instead, a so-called Moriyaesque $\lambda$-correction  \cite{RMPVertex}. It essentially adds a mass to the paramagnons (dampens the antiferromagnetic spin fluctuations). This mass is fixed by a sum-rule for the susceptibility, and automatically warranties the correct high-frequency asymptotics of the self-energy. This $\lambda$-correction has been introduced to mimic the self-consistency and represents   a considerable simplification of the calculations. For a further-going discussion see \cite{RMPVertex,Kitatani2019}.

The advantage of the ladder D$\mathrm \Gamma$A is that --as long as we do not couple the ladders through the
parquet equations-- the ladder only depends on a single frequency-momentum $q$
instead of three  ($q$,$k$,$k^\prime$). Hence numerically much lower temperatures, larger frequency grids and  finer momentum grids are feasible.
Also realistic multi-orbital {\em ab initio} D$\mathrm \Gamma$A calculations are possible with the ladder D$\mathrm \Gamma$A version \cite{Galler2016}.
The results presented in the following have been obtained by ladder D$\mathrm \Gamma$A with a Moriyaesque $\lambda$-correction.

\section{Hubbard model, cuprates and nickelates}
\label{Sec:model}\label{Sec:HM}
In this Chapter, we consider the one-band Hubbard model in two or three-dimensions:
\begin{equation}
\mathcal{H} = -\sum_{ij,\sigma}  t_{ij} c_{i\sigma}^{\dagger}c_{j\sigma}^{\phantom{\dagger}} + U\sum_{i}n_{i\uparrow}^{\phantom{\dagger}}n_{i\downarrow}^{\phantom{\dagger}}.
\label{eq:Hubbard} \index{Hubbard model}
\end{equation}
It consists of two terms: (i) a  hopping term $t_{ij}$ between sites $i$ and $j$, which we restrict in the following to a nearest neighbor $t$, next-nearest neighbor $t'$ and next-next-nearest neighbor $t''$;  and (ii) a   local Coulomb repulsion $U$.
Here $c_{i\sigma}^{\dagger}$ ($c_{i\sigma}^{\phantom{\dagger}}$) creates (annihilates) an electron on site $i$ with spin $\sigma$ in second quantization and 
 $n_{i\sigma}^{\phantom{\dagger}}=c_{i\sigma}^{\dagger}c_{i\sigma}^{\phantom{\dagger}}$.

The Hubbard model is the quintessential model for strongly correlated electron systems, similar as the Ising model for statistical physics or the Drosophila fly for genetics. For $t_{ij}=0$, we have just a collection of atoms and a spectra with peaks at $\pm U/2$ for half-filling. For $U=0$, there is no interaction and we can solve the tight-binding Hamiltonian by Fourier transforming $t_{ij}$ to $\varepsilon_{\mathbf k}$ in momentum space. Then, just occupying all single-particle states up to the Fermi energy gives the ground state.  But, if we  switch on $U$ the electrons become correlated, the expectation value  $\langle n_{i\sigma} n_{j\sigma'} \rangle  \neq \langle n_{i\sigma} \rangle \langle n_{j\sigma'} \rangle$. In particular local double occupations  $\langle n_{i\uparrow} n_{i\downarrow} \rangle$ are heavily reduced compared to the non-interacting, uncorrelated value.

While a screened interaction can --to a good approximation-- often be replaced by the purely local interaction $U$, materials require typically the consideration of more bands than the single-band Hubbard model. This is even the case when we restrict ourselves to the low-energy orbitals around the Fermi energy. An important exception are cuprate and nickelate superconductors. \index{cuprates} \index{nickelates} 

\begin{figure}[t!]
 \centering
 \includegraphics[width=0.95\textwidth]{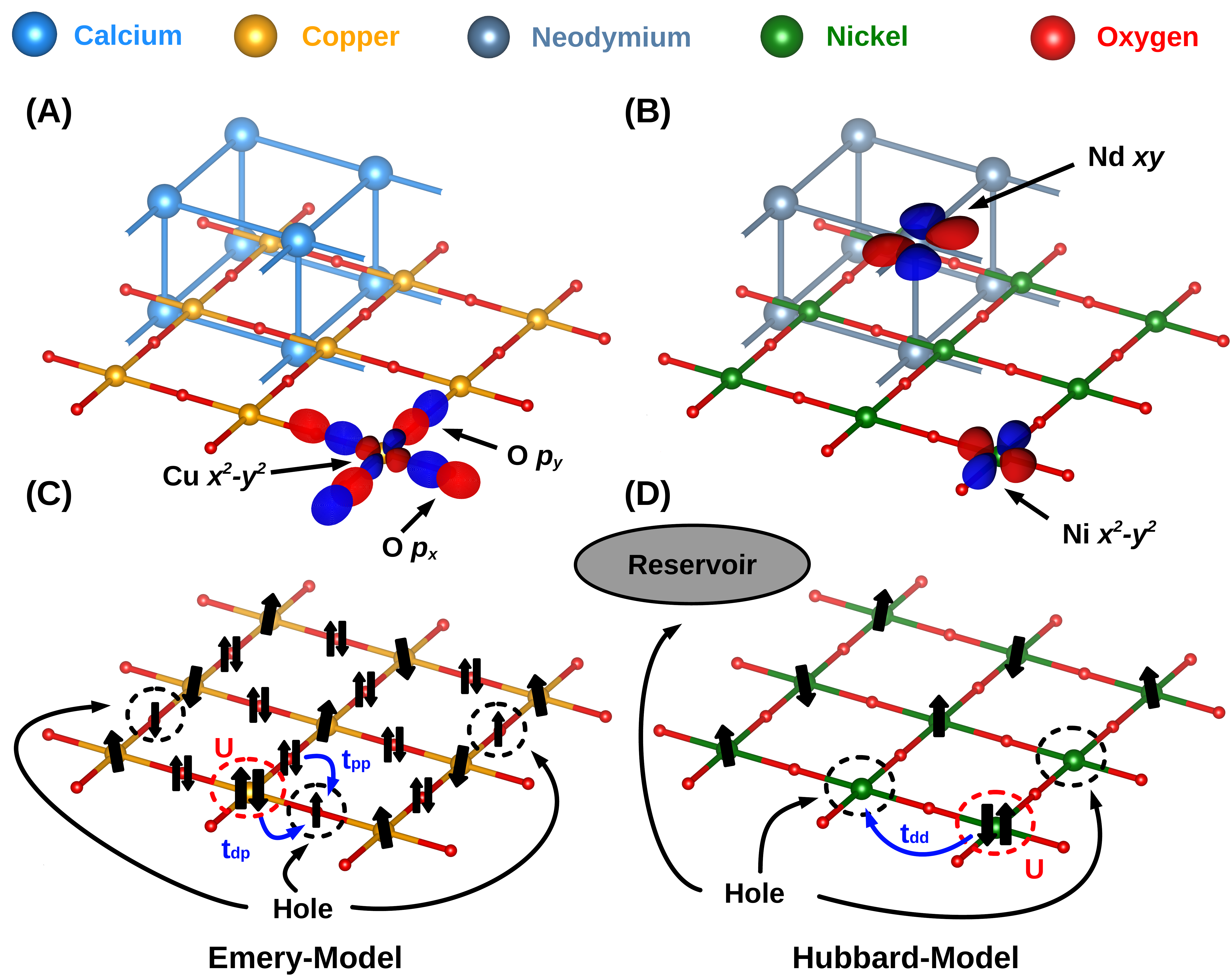}
  \caption{Crystal structure of (A) a characteristic cuprate, CaCuO$_2$,  and (B) the nickelate NdNiO$_2$.
The  Cu(Ni)O$_2$ layers are separated by Ca(Nd) spacing layers in this "infinite-layer" structure. Also shown are the relevant low-energy orbitals in both cases. 
(C) Cuprates are charge transfer insulators so that the oxygen orbitals need to be included for an accurate microscopic description.
The simplest such model is the {\bf Emery model}
with copper  3$d_{x^2-y^2}$, oxygen $p_x$ and $p_y$ orbitals.
As indicated there  is a hopping $t_{pd}$ between Cu and O sites, and
$t_{pp}$ between O sites; double occupations are suppressed by the interaction $U$ on the Cu sites just as in the Hubbard model. (D) For nickelates  we have a  Ni-3$d_{x^2-y^2}$-band and a $A$-pocket derived from {the} Nd 5$d_{xy}$ {band} crossing the Fermi energy. Both are largely decoupled but we need to calculate, e.g., by DFT+DMFT, how many holes go into the 
Ni-3$d_{x^2-y^2}$-band  {\bf Hubbard model}
and how many go into the $A$-pocket {\bf reservoir}. From \cite{Held2022}. \label{Fig:HM}
 }
\end{figure}

The arguably simplest cuprate and nickelate crystal structure is the ``infinite-layer''\footnote{The two distinct layers displayed in Fig.~\ref{Fig:HM} are repeated on top of each other ad infinitum.}  structure displayed in Fig.~\ref{Fig:HM} (A) and (B).
Since the valence of the spacer cations is Ca$^{+2}$ and  Nd(La)$^{+3}$, the formal oxidation state is  Cu$^{+2}$ and  Ni$^{+1}$, respectively, so that both cuprates and nickelates are in a formal $3d^9$ configuration.
After such basic chemistry considerations,
the first step to get an idea of the relevant orbitals is doing a density functional theory (DFT) \index{density functional theory} calculation.

\begin{figure}[t!]
 \centering
 \includegraphics[width=0.65\textwidth]{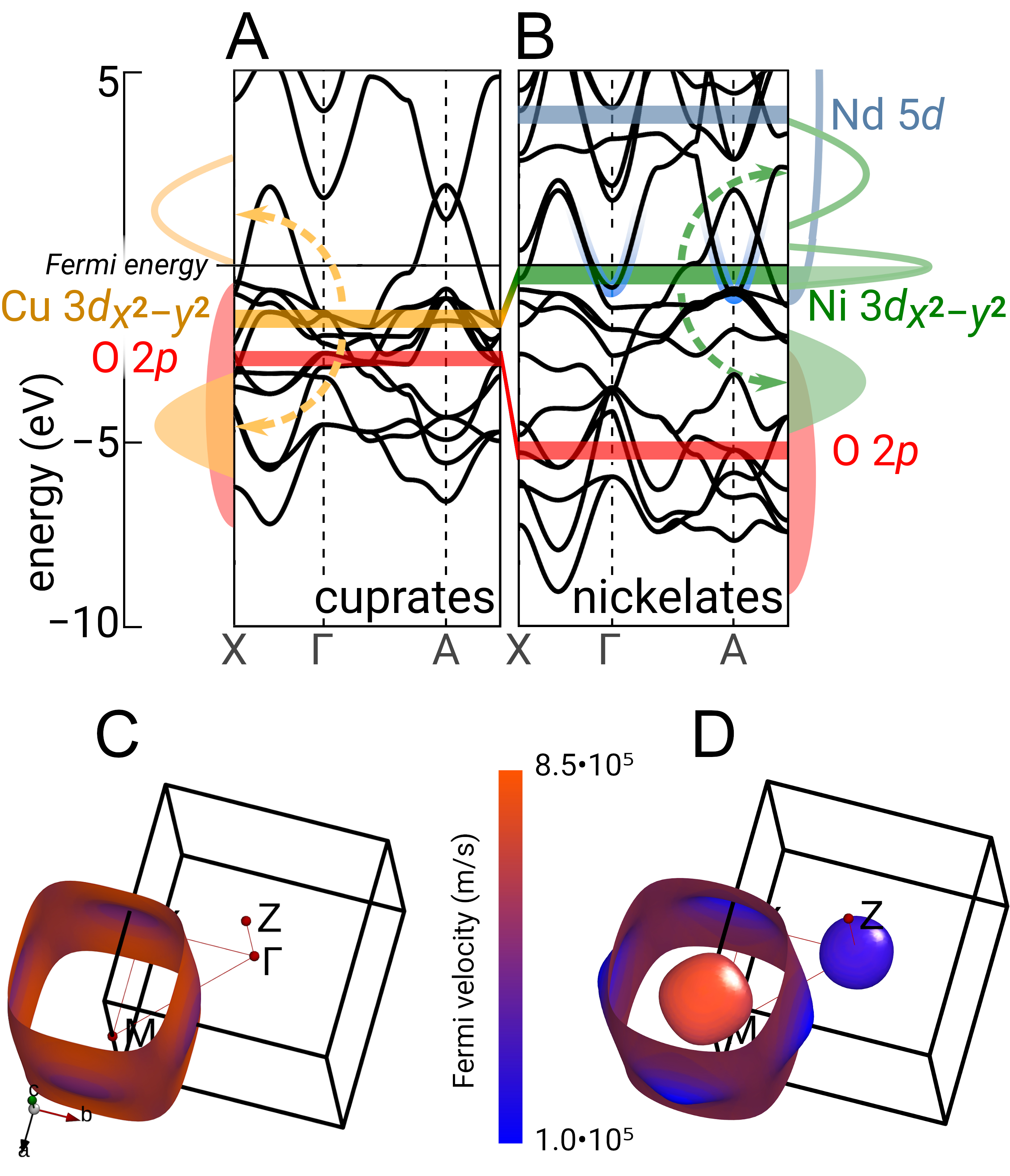}
 \vspace{.3cm}
 
   \caption{(A,B) Electronic structures of (A) CaCuO$_2$  and (B) LaNiO$_2$ as calculated in DFT.  
The side panels indicate the effect of electronic correlations and the color bars the  center of energy of the most important  bands.
(C) and (D) DFT Fermi surface corresponding to (A) and (B), respectively. From \cite{Held2022}.
  \label{Fig:cuprates}
 }
\end{figure}

 Fig.~\ref{Fig:cuprates} shows that for the cuprates a single $3d_{x^2-y^2}$-orbital is crossing the Fermi level. It gives rise to the single hole-like Fermi surface sheet displayed in the lower left panel of 
 Fig.~\ref{Fig:cuprates}. This Cu $3d_{x^2-y^2}$-band  and Fermi surface can well be described with proper hopping parameters $t$, $t'$ and $t''$.
 But now we need to include the effect of electronic correlations. These are indicated by the dashed arrows and  the side panels of  Fig.~\ref{Fig:cuprates} (A): the  Cu $3d_{x^2-y^2}$-band splits into an upper and lower Hubbard band. Since the oxygen orbitals are just a few eV below the Fermi energy, these oxygen orbitals end up above the lower Hubbard band. As a consequence we have a charge transfer insulator according to the scheme of Zaanen-Sawatzky-Allen and not a Mott insulator, which one might have expected from the splitting into Hubbard bands.
 That is, if we dope cuprates, and we need to do so to have a superconductor, the holes go into the oxygen bands. Hence, in the case of cuprates, an Emery model description which incorporates  the copper $3d_{x^2-y^2}$-band and oxygen $p_x$ and $p_y$ bands as visualized in  Fig.~\ref{Fig:HM} (C) is the most appropriate low-energy model. Nonetheless, the majority of theoretical papers studying the cuprates use the single-band Hubbard model. This is to some extend justified by the fact that oxygen-hole spin and copper spin form a Zhang-Rice singlet, which can be described effectively  by the Hubbard model. Also in experiment, there is a single Fermi surface as in   Fig.~\ref{Fig:cuprates} (C) of mixed oxygen and copper character.

In case of the nickelates, these oxygen bands are instead at a considerably lower energy. Hence the lower Hubbard band is now above the oxygen band and we would have a Mott insulator -- if it was not for the Nd(La) $4d$-bands.\footnote{We here show LaNiO$_2$ since it is slightly easier to calculate than NdNiO$_2$ in DFT because it has no $4f$-electrons. Experimentally,  La$_{1-x}$Sr(Ca)$_x$NiO$_2$ is also a nickelate superconductor with a  $T_c$ comparable to NdNiO$_2$.}  These Nd(La) $4d$ bands are closer to the Ni $3d$-bands than the unoccupied Ca  and  Cu $s$ bands are to the Cu bands. Hence they 
briefly cross the Fermi level  around the $\Gamma$ and $A$ momentum, which leads to the formation of pockets around these momenta, see  Fig.~\ref{Fig:cuprates} (D). The  charge  transferred to the Nd(La) pockets also leads to a self-doping of the Ni  $3d_{x^2-y^2}$-band, even for the parent compound Nd(La)NiO$_2$ which has about 0.05 holes per Ni. In such a situation, a quasi-particle peak develops as indicated in the side panel of  Fig.~\ref{Fig:cuprates} (B) with a quasiparticle mass enhancement $m^*/m=1/Z$ calculated to be about five \cite{Kitatani2020}.
A further correlation effect is that the $\Gamma$-pocket is shifted above the Fermi level, at least for larger Sr-doping and for LaNiO$_2$.

Altogether this leaves us with two relevant bands as displayed in  Fig.~\ref{Fig:HM} (B): the Ni  $3d_{x^2-y^2}$-band and the Nd(La)-derived $A$-pocket.
Both bands do not hybridize and can hence, to a first approximation, be considered as decoupled. The expectation is that the more strongly correlated Ni $3d_{x^2-y^2}$-band is responsible for the superconductivity. Using Occam's razor, i.e., if we try to identify the most simple model, we end up with a one-band Hubbard model for the Ni  $3d_{x^2-y^2}$-band and a decoupled reservoir ($A$-pocket) that must be
taken into account for translating the Sr-doping of  Nd(La)$_{1-x}$Sr$_x$NiO$_2$ into the actual hole doping of this Hubbard model. Otherwise the $A$-pocket is merely a passive bystander.
This simple model is illustrated in Fig.~\ref{Fig:HM} (D).

The  hopping parameters of the nickelate Hubbard model can be obtained from a Wannier function projection onto the Ni  $3d_{x^2-y^2}$-band: $t=0.395\,$eV, $t'/t=-0.25$, $t''/t=0.12$ \cite{Kitatani2020}. Further, the interaction strength $U$ can be calculated by constrained random phase approximation (cRPA, see Chapter ``The GW+EDMFT method''  by F. Aryasetiawan \cite{Pavarini2022}). Considering the fact that $U$  is frequency($\omega$)-dependent in cRPA and --within the relevant energy range-- on average slightly larger than $U(\omega=0)$, one obtains $U\approx 8t$ from cRPA. For further details see \cite{Kitatani2020}.

\begin{figure}[t!]
 \centering
 \includegraphics[width=0.925\textwidth]{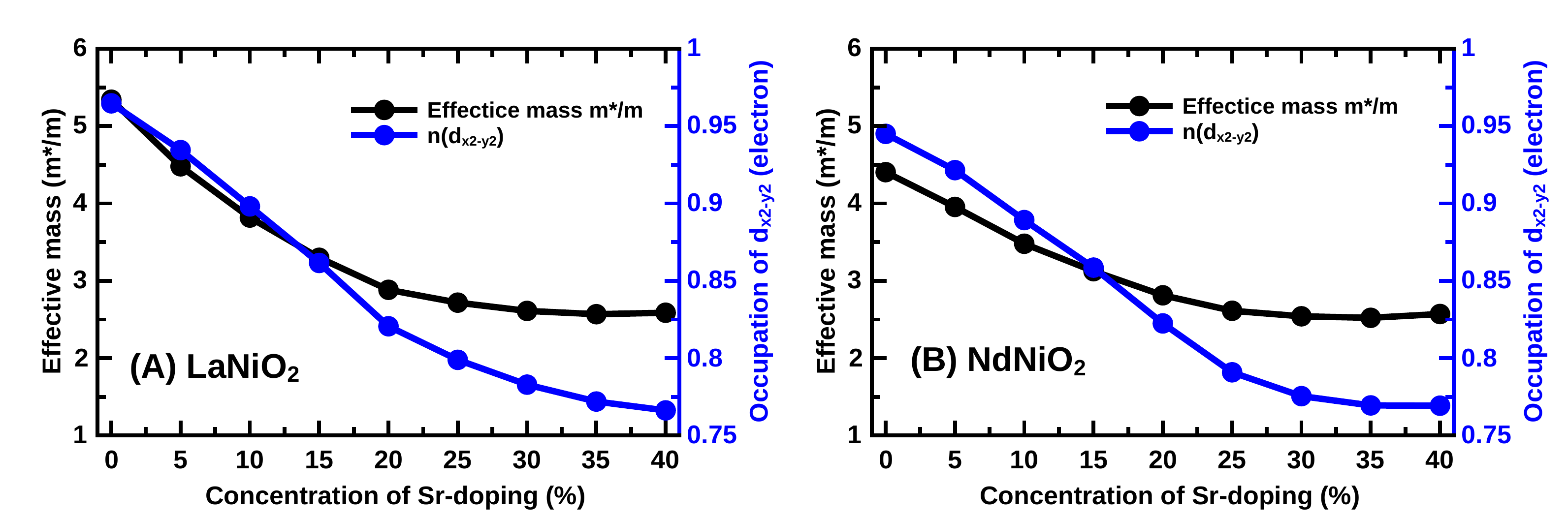}
 \caption{Translation of Sr-doping to the occupation of the Ni $3d_{x^2-y^2}$-band (blue line; right $y$-axis)
and corresponding effective mass enhancement $m^*/m$ (black line; left $y$-axis) as calculated by DFT+DMFT for (A) La$_{1-x}$Sr$_x$NiO$_2$ and (B) Nd$_{1-x}$Sr$_x$NiO$_2$.
   From \cite{Kitatani2020} (Supplemental Material).
  \label{Fig:doping}
 }
\end{figure}

The translation from Sr-doping to the occupation of the $3d_{x^2-y^2}$-band has been calculated in DFT+DMFT\footnote{For the DFT+DMFT method, see the Chapter "LDA+DMFT for strongly correlated materials" by A.~Lichtenstein \cite{Pavarini2022}} including all five Nd and all five Ni $d$ bands in DMFT.  Fig.~\ref{Fig:doping} (blue curve) shows the results: One sees that roughly 50\% of the holes (there is one hole per Sr) go into the  $3d_{x^2-y^2}$-band and the remaining 50\% go into the pockets. For larger Sr-dopings, at around 25\%, the curve flattens because here the Ni $3d_{3z^2-r^2}$ orbital approaches the Fermi level and accommodates holes as well.
The one-band Hubbard model DMFT calculation gives a very similar spectrum as the full DFT+DMFT calculations with 5+5 Nd+Ni orbitals. Also the effective mass plotted in Fig.~\ref{Fig:doping} (black curve) agrees  \cite{Kitatani2020}.
Altogether this hints that the simple Hubbard model is a good approximation for nickelates; experimental results are also consistent with this picture so-far.

\section{Spin fluctuations}
\label{Sec:spin fluct}
\index{spin fluctuations}

\begin{figure}[t!]
  \begin{center}
    \vspace{.2cm}
    
    \includegraphics[width=.85\textwidth]{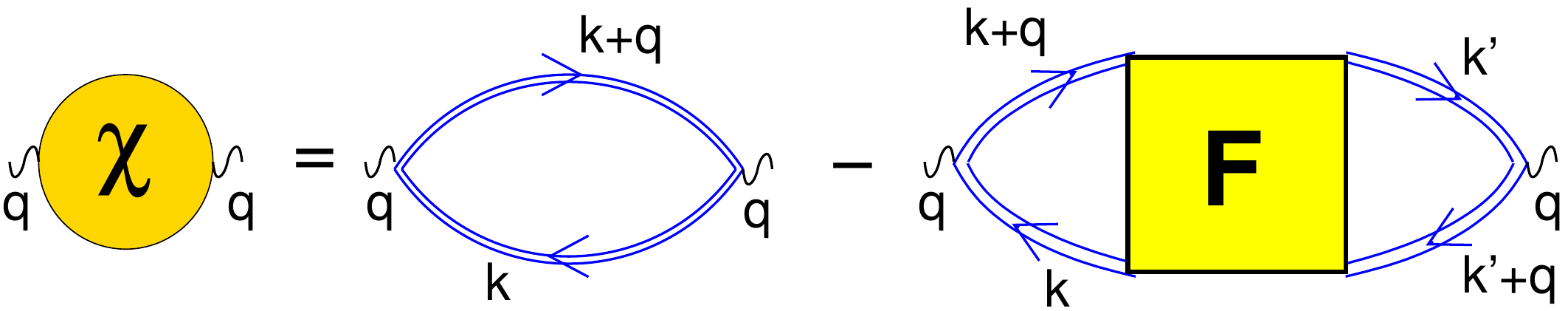}
  \end{center}
  \caption{The susceptibility $\chi$ consists of the bare bubble $\chi^0$ (first term on the right hand side) and vertex corrections that can be calculated from the full vertex $F$.\index{susceptibility}\label{Fig:sus}}
\end{figure}

Spin fluctuations as visualized in  Fig.~\ref{Fig:ladderDGA} (a,b) enter the full vertex $F$ but subsequently also the susceptibility which is connected to $F$ as follows (Fig.\ref{Fig:sus}):
\begin{equation}
  \chi_{r,{\mathbf q} \omega} =  \underbrace{
    \sum_{{\mathbf  k} \nu} G_{{\mathbf k} \nu}    G_{{(\mathbf k +\mathbf q)} (\nu +\omega)}
  }_{\equiv \chi^0_{{\mathbf q} \omega}}
-   \sum_{{\mathbf  k} {\mathbf  k{\prime}} \nu \nu^{\prime}}
G_{{\mathbf k} \nu}    G_{{(\mathbf k +\mathbf q)} (\nu +\omega)} F_{r, \mathbf k +\mathbf k^{\prime}  \mathbf q}^{\nu \nu^{\prime}  \omega}  G_{{\mathbf k^{\prime}} \nu}    G_{{(\mathbf k^{\prime} +\mathbf q)} (\nu +\omega)}
\label{Eq:sus}
  \end{equation}
for $r=c/s$ i.e.,  the spin  and charge susceptibility, respectively. As before, we restrict ourselves to the paramagnetic phase for simplicity. 
The first term on the right hand side of Eq.~(\ref{Eq:sus}) is just the bare bubble contribution $\chi^0$, obtained directly  from two (interacting) Green's function; the second term are vertex corrections calculated form $F$. The minus sign is a matter of definition of $F$. When $F$ is calculated in RPA as in Fig.~\ref{Fig:ladderDGA} (a),  we have $\Gamma_r=\pm U$ for $r=c/s$ and obtain a geometric sum, which eventually yields
\begin{equation}
  \chi^{\mathrm RPA}_{r,{\mathbf q} \omega} =   \frac{\chi^0_{{\mathbf q} \omega} }{1 \pm U  \chi^0_{{\mathbf q} \omega}}
\label{Eq:sus2}
\end{equation}
where the +/- is for $r=c/s$.

For certain $\omega$'s and $\mathbf q$'s
Eq.~(\ref{Eq:sus2}) [and  Eq.~(\ref{Eq:sus})] develops poles. We call these bosonic excitations (para)magnons. \index{magnon} The $\omega$-$\mathbf q$ energy-momentum dispersion relation of these quasiparticles follows the position of the poles.

The minus sign in Eq.~(\ref{Eq:sus2}) already indicates that spin fluctuations are generally stronger in a Hubbard model. This can change if additional non-local interactions $V$ are included. These trigger a transition from an antiferromagnetic spin  to a charge density wave order for ${\cal Z} V>U$ in mean field (${\cal Z}$:  number of neighbors; $V$: nearest neighbor Coulomb repulsion). Which magnetic order dominates in RPA, solely depends on the $\mathbf q$ for which $\chi^0_{{\mathbf q} \, \omega=0}$ is strongest. Close to half-filling, we often have Fermi surfaces where for a ${\mathbf k}$ on the Fermi surface ($\nu=0$)
also ${\mathbf k} +{\mathbf q}=(\pi,\pi,\ldots)$ is at or close to the Fermi surface.
Then   both $G_{{\mathbf k} \nu=0}$  and $G_{{(\mathbf k  + (\pi,\pi,\ldots))} (\nu +\omega)=0}$ are large in   Eq.~(\ref{Eq:sus}), and the antiferromagnetic susceptibility 
$\chi^0_{{\mathbf q}=  (\pi,\pi,\ldots) \omega=0}$ is maximal. Of course, this is just the weak coupling picture. For larger Coulomb interactions we form large magnetic moments  and the change of physics is reflected in vertex $F$ corrections  beyond RPA.

\begin{figure}[t!]
  \begin{center}
    \includegraphics[width=.47\textwidth,trim={20.18cm 17.915cm 0 0 },clip=true]{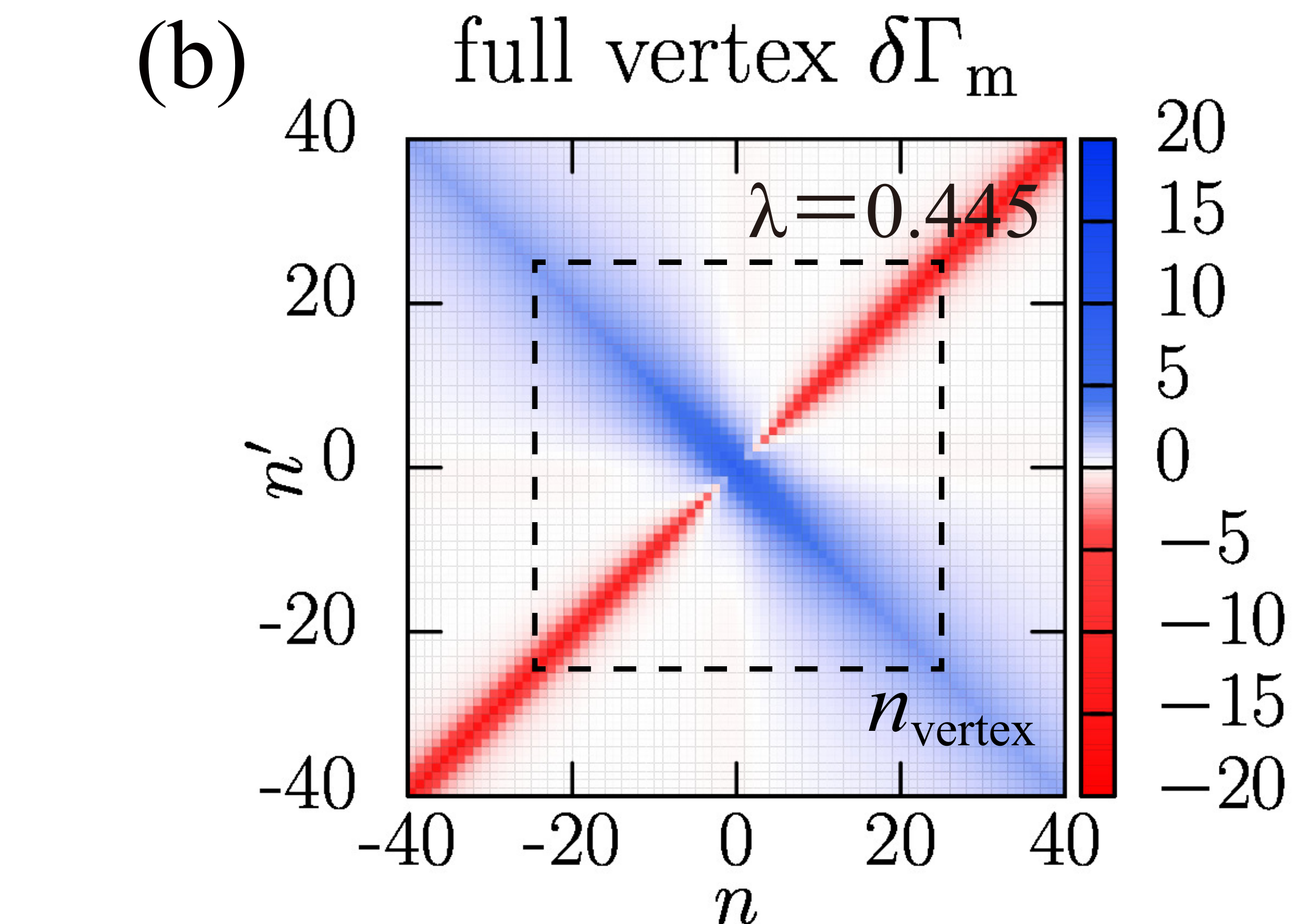}
    \hfill 
   \includegraphics[width=.425\textwidth,trim={22.78cm 0 0 18.38cm},clip=true]{FigSuppGamma.pdf} 
  \end{center}
  \caption{Left: Difference of the local $ph$ vertex in the magnetic channel from $-U$, i.e., $\delta \Gamma_m=\Gamma_{s}-(-U)$ vs.\ the two fermionic Matsubara frequencies $\nu_n=i (2n+1)\pi T$ and  $\nu'_{n'}$ at zero bosonic frequency $\omega=0$. At small frequencies,  $\Gamma_{s,ph}$ and associated spin fluctuations are  suppressed.
    Right: This suppression also leads to a suppression of the non-local
    $pp$ vertex $\Gamma_{pp}$, i.e., the superconducting pairing glue, at small frequencies. It is obtained from antiferromagnetic spin fluctuations (i.e., from the local $\Gamma_{s}$ corresponding  to the left panel and the Bethe-Salpeter equation, see Sec.~\ref{Sec:SC}). Right inset:  $\Gamma_{pp}$ as calculated in RPA.
From \cite{Kitatani2019} where further details and parameters can be found. \label{Fig:Gammasuppress} \index{vertex}}
\end{figure}

More recently it could be demonstrated 
\cite{Kitatani2019} that the local vertex $\Gamma_{r=m}$ is suppressed compared to the RPA $\Gamma_s=-U$ value. This is shown in
Fig.~\ref{Fig:Gammasuppress} (left). This suppression, in particular that at the relevant small frequencies, leads to reduced spin fluctuations. Consequently, the superconducting pairing as calculated along the line of  Fig.~\ref{Fig:ladderDGA} (b,c) is suppressed as well, see
Fig.~\ref{Fig:Gammasuppress} (right). Its origin are local particle-particle excitations that enter $\Gamma_{s}$ and reduce it along the side diagonal frequencies $\nu=-\nu'$.

This reduction of the D$\mathrm \Gamma$A spin susceptibility is key for a good description of antiferromagnetic spin fluctuations which are grossly overestimated in RPA for somewhat larger Coulomb interactions $U$. In fact, it was shown that qualitatively and quantitatively the D$\mathrm \Gamma$A susceptibility excellently agrees with other recent numerical calculations \cite{Schaefer2021}.

\begin{figure}[t!]
  \begin{center}
    \vspace{.2cm}
    
    \includegraphics[width=.95\textwidth,trim={ 0 8.818cm 0 10.9818cm },clip=true]{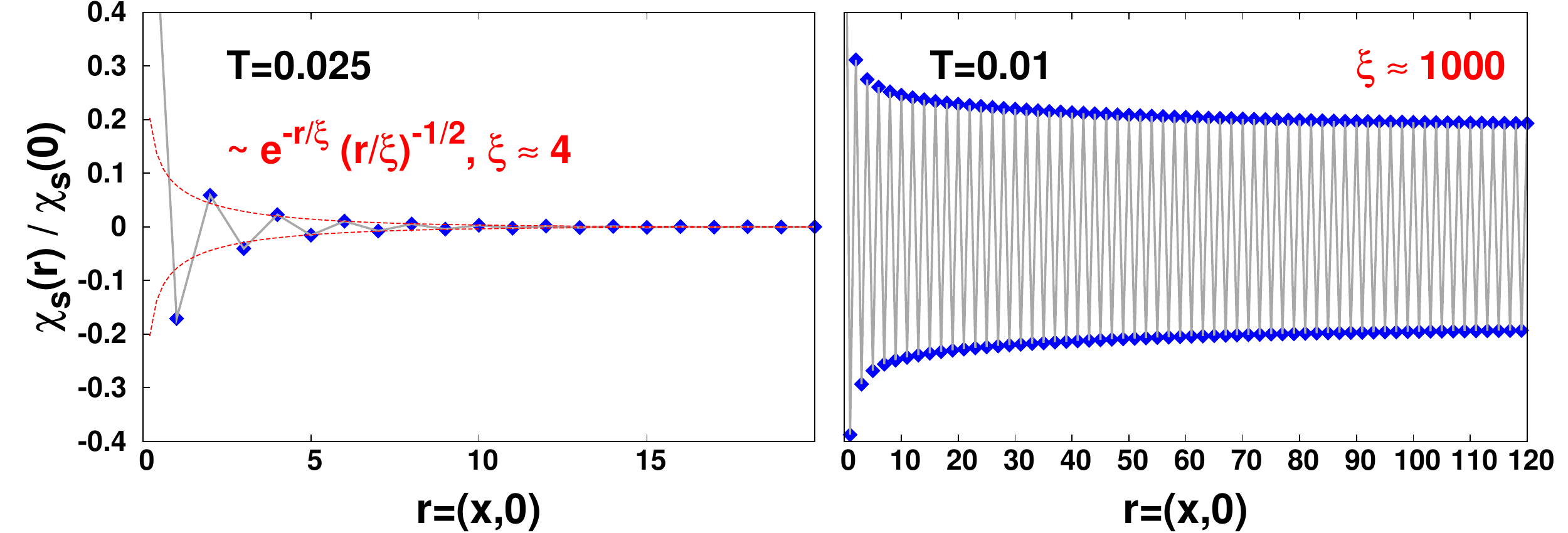}
  \end{center}
  \caption{Antiferromagnetic correlation function $\chi_{s, {\mathbf r}}$ normalized to its $\mathbf r=0$ value  vs.\ distance $\mathbf r$ for the two dimensional Hubbard model along the $x$-direction ($t'=t''=0$, $U=0.5 (4t)$;  all energies are here in units of $4t$). The left and right panel show two different temperatures, $T=0.025 (4t)$ (left) and  $T=0.025 (4t)$ (right), where largely different correlation lengths ($\xi$) are observed.  From \cite{Schaefer2015-2}.\label{Fig:corr} \index{spin fluctuations}}
\end{figure}

From the susceptibility one can 
obtain the correlation length $\xi$. Fourier-transformed to real space ${\mathbf r}$ and time,
the (equal-time) magnetic susceptibility behaves as
\begin{equation}
  \chi_{s {\mathbf r}} = \langle {\mathbf S} (\mathbf r) {\mathbf S} ({\mathbf 0}) \rangle \sim \left(\frac{\scriptstyle |{\mathbf r}|}{\scriptstyle \xi}\right)^{-1/2} e^{-\frac{\scriptstyle |{\mathbf r|}}{\scriptstyle  \xi}}
\end{equation}
at large distances $\mathbf r$. Here, ${\mathbf S} (\mathbf r)$ denotes the spin operator of the electrons at position (lattice site) $\mathbf r$.
Fig.~\ref{Fig:corr} shows this spin-spin correlations or susceptibility of the Hubbard model. Clearly, an alternating (antiferromagnetic) correlation is visible. At high temperatures (left panel), the correlation length is short and correlations quickly decay. In two dimensions we get however an exponential increase of the correlation length with $1/T$ in D$\mathrm \Gamma$A  \cite{Schaefer2015-2} which is the reason behind the very large correlation length of  Fig.~\ref{Fig:corr}  (right panel).
While a correlation length of $\xi=4$ lattice sites (left panel) can still be covered in cluster extensions of DMFT and lattice quantum Monte Carlo methods, the rapidly  increasing correlation length at lower temperatures quickly puts a numerical limit to such cluster approaches.

At a three dimensional phase transition, the susceptibility diverges with critical exponent $\nu$:  $\xi\sim (T-T_c)^{-\nu}$, and similarly $\chi_{r, {\mathbf Q} \omega=0} \sim (T-T_c)^{-\gamma}$ with a critical exponent $\gamma$. These critical exponents could be calculated in D$\mathrm \Gamma$A and DF for the first time for correlated electronic models, a topic that has been covered in a preceding J\"ulich Autumn School  \cite{Held2018}.

In practice, the correlation length is not calculated from fitting  $\chi_{s {\mathbf r}}$. Instead one fits it Fourier-transform $\chi_{s, {\mathbf q}\omega=0}$ in momentum space. For large correlation lengths and in the vicinity of its maximum at the dominating wave vector $\mathbf Q$,\footnote{For example, ${\mathbf Q}=(\pi,\pi)$ for the two-dimensional Hubbard model at half-filling.} the susceptibility  is of the Ornstein-Zernike form:
\begin{equation}
  \label{eq:OZ}
  \chi_{s, {\mathbf q}\,  \omega=0} \sim \frac{1}{({\mathbf q}-{\mathbf Q})^2+\xi^{-2}}.
  \end{equation}
That is, the inverse correlation length $\xi^{-1}$ corresponds to the width of the susceptibility around its peak at ${\mathbf q}={\mathbf Q}$.

\begin{figure}[t]
   \centering \includegraphics[width=.7\columnwidth,angle=0]{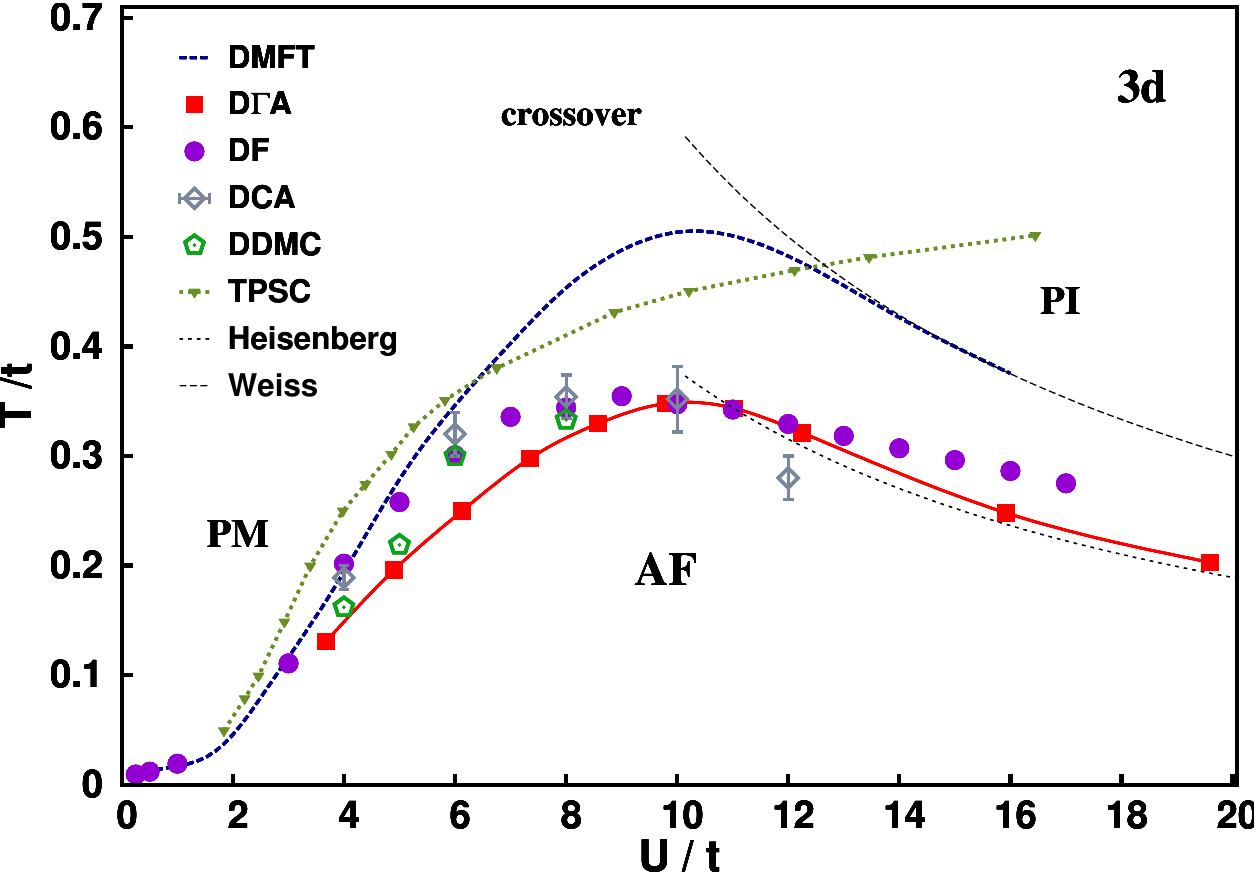} 
   \caption{Phase diagram of the half-filled three-dimensional Hubbard model on a cubic lattice with nearest-neighbor hopping $t$ comparing   various methods
     for the  phase transition from the paramagnetic metallic (PM) and insulating (PI) phase to the antiferromagnetic insulating phase (AF). From \cite{RMPVertex} where further details and the abbreviation of the various methods can be found. \index{antiferromagnetism}}
  \label{fig:pd3d}
\end{figure}

In two dimensions, the Mermin-Wagner theorem is fulfilled and there is no long range antiferromagnetic order for finite temperature in the D$\mathrm \Gamma$A and the wit diagrammatic extension of DMFT \cite{RMPVertex}. In three-dimensions the antiferromagnetic phase transition temperature $T_c$ is reduced compared to DMFT, as shown in Fig.~\ref{fig:pd3d}. Also the critical behavior in the vicinity of the phase transition is not any longer of mean-field type as in DMFT. Instead the critical exponents well agree\footnote{within the numerical error bars in the studied critical temperature regime} with those of the Heisenberg model \cite{RMPVertex,Held2018} for half-filling, as to be expected from universality.

\section{Pseudogap}
\label{Sec:PG}
\index{pseudogap}

\begin{figure}[t!]
  \begin{center}
    \includegraphics[width=0.95\textwidth]{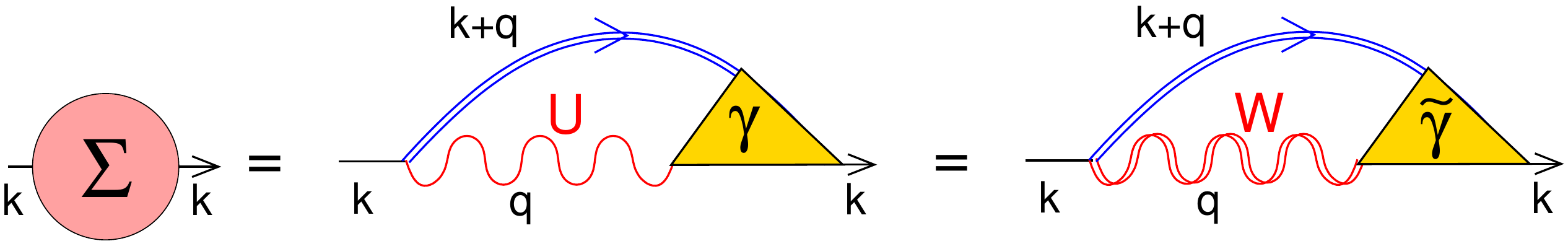}
  \end{center}
 \caption{Schwinger-Dyson equation to calculate the self-energy $\Sigma$  rewritten in terms of $U$ and $\gamma$ or $W$ and $\tilde\gamma$.\label{Fig:SDE2} \index{Schwinger-Dyson equation}}
\end{figure}

Besides the susceptibility, spin fluctuations also impact the electronic self-energy, as already indicated in Fig.~\ref{Fig:selfenergy}.
The effect $F$  on the self-energy $\Sigma$ can be directly calculated via the  Schwinger-Dyson Eq.~(\ref{eq:SD}) [Fig.~\ref{Fig:SDE}]. Physically, this corresponds to electron-(para)magnon scattering which can reduce the lifetimes of the quasiparticles.
An alternative way to write the Schwinger-Dyson equation and to express the fermion-boson interaction is shown in Fig.~\ref{Fig:SDE2}. Here, simply the $F$ and two Green's functions from Fig.~\ref{Fig:SDE} have been combined to  $\gamma$, i.e., mathematically
\begin{equation}
\gamma_{r,{\mathbf k} {\mathbf q}}^{\nu\omega}=  \sum_{{\mathbf k^\prime} \nu^\prime} F_{{r}, {\mathbf k \mathbf k^{\prime}\mathbf q}} G_{{\mathbf  k^{\prime}}\nu^{\prime}}G_{({\mathbf k^{\prime}+\mathbf q})(\nu^{\prime}+\omega)} 
\end{equation}
so that the Schwinger-Dyson equation becomes
\begin{align}\label{eq:SD2}
\Sigma_{{\mathbf k}\nu}&= \frac{U n}{2}-\frac{U}{2}\sum_{{\mathbf q} \, \omega}\left[\gamma_{c,{\mathbf k} {\mathbf q}}^{\nu\omega} - \gamma_{s,{\mathbf k} {\mathbf q}}^{\nu\omega} \right] G_{({\mathbf k+ \mathbf q})(\nu+\omega)} \; .
\end{align}
Now, we can rewrite this further by merging $U$-reducible diagrams into an effective interaction $W_r$, in the spirit of Hedin's $GW$ method 
(we will not go into further details here and refer the reader to the Chapter "The GW+EDMFT method'' by F.~Aryasetiawan \cite{Pavarini2022},  and to \cite{Krien2019}). Beyond $GW$, here also  spin fluctuations are included in $W_s$. This is displayed in Fig.~\ref{Fig:SDE2} (right).  The remaining $U$-irreducible 
$\tilde \gamma$ can be interpreted as the fermion-boson interaction and $W$   as a  boson propagator. In the case of $GW$ these bosons are plasmons (charge fluctuations), while for the one-band Hubbard model (para)magnons (spin fluctuations) dominate.
Such a reformulation of the Feynman diagrams is also the  idea of the single boson exchange (SBE) approach  \cite{Krien2019}, which has been used recently to rewrite the DF and D$\mathrm \Gamma$A parquet equations. Much of the physics is already contained in the boson propagators $W$ so that the remaining fully irreducible parts of the SBE parquet formalism decay much faster in frequency and momentum.

Particularly strong spin fluctuations, as they occur for the Hubbard model in two dimensions, can give rise to a physical phenomenon called pseudogap. Theoretically, this pseudogap has been established in numerical simulations and in weak coupling calculations. Experimentally, it has been observed for cuprate superconductors, both in the spectrum as well as in transport measurements. This kind of physics is naturally included in D$\mathrm \Gamma$A and other diagrammatic extensions of DMFT.

\begin{figure}[t!]
  \begin{center}
    \includegraphics[width=.4\textwidth]{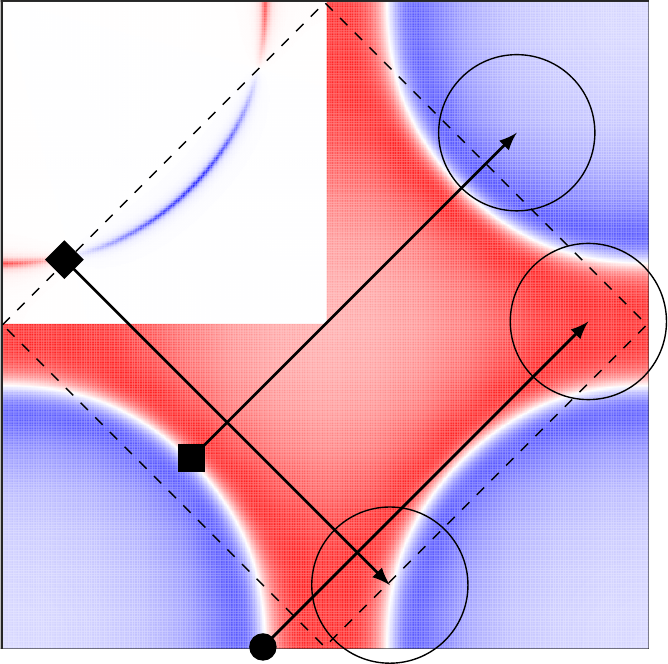}
  \end{center}
  \caption{Typical Fermi surface of the $t$-$t'$-$t''$ Hubbard model. The blue and red colors denote the occupied and unoccupied part of the Brillouin zone, respectively. 
    The circle (PG), square (ARC) and diamond (HS) denote specific momenta ${\mathbf k}$ that are  discussed in the text. The corresponding momenta ${\mathbf k} +{(\pm \pi,\pm \pi)}$ connected to these by  the antiferromagnetic wave vector ${\mathbf Q}=(\pi,\pi)$ are also plotted, as circles with typical  inverse correlation lengths $\xi^{-1}$ as radius. The upper left quadrant shows the parts of the Fermi surface that are "heated" (red) and "cooled" (blue), i.e., become less and more coherent because of  ${\rm Im} \tilde \gamma$ (see text).  From \cite{Krien2021}. \label{Fig:FS}  }
\end{figure}

\begin{figure}[t!]
  \begin{center}
    \includegraphics[width=.85\textwidth,trim={ .051cm 0 0.025cm .29cm},clip=true]{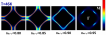}
  \end{center}
  \caption{D$\Gamma$A spectrum at the Fermi  energy  for different dopings  of nickelate superconductors throughout the Brillouin zone
    (i.e,  $x$- and $y$-axis are the $k_x$ and $k_y$ momentum ranging from $-\pi$ to $\pi$, and the false color show the spectral function at the Fermi energy $\nu=0$). From \cite{Kitatani2020}. \label{Fig:PGDGA} \index{nickelates}\index{pseudogap} }
\end{figure}

Fig.~\ref{Fig:FS} shows the Fermi surface of the two-dimensional Hubbard model for typical hopping parameters of Cu(Ni) $3d_{x^2-y ^2}$ orbitals in cuprates (nickelates). If we want to calculate the self-energy  for a momentum ${\mathbf k}$ on the Fermi surface, it will be effected by spin fluctuations. These effects can be calculated via  Eq.~(\ref{eq:SD2}) [Fig.~\ref{Fig:SDE2}].
In  Fig.~\ref{Fig:FS} three particular points of the Fermi  surface are displayed:
The antinodal  momentum on the Fermi surface (PG) is close to  $(\pi,0$). Here, the pseudogap first opens in experiment and in numerical calculations for the Hubbard model at strong coupling. The nodal momentum (ARC) along the diagonal  where an ARC-like part of the Fermi surface survives after the opening of the pseudogap around PG [see Fig.~\ref{Fig:PGDGA} (rightmost panel) below]. Finally, the hot spot (HS) which is the point of the Fermi surface where ${\mathbf k}+(\pi,\pi)$  lies on the Fermi surface as well. At weak coupling, the pseudogap opens first at the HS.

The spin-fermion coupling of Fig.~\ref{Fig:SDE2} leads to an additional contribution to the self-energy  on top of the local DMFT self-energy.
In the Bethe-Salpeter equation Eq.~(\ref{eq:SD}) or  equivalently Eq.~(\ref{eq:SD2}), the vertex $F$ respectively $\gamma$ -- or in  Fig.~\ref{Fig:SDE2} (right )$W$-- are dominated by spin-fluctuations for the two-dimensional Hubbard model. These spin fluctuations are strongly peaked around the antiferromagnetic weave vector
${\mathbf Q}=(\pi,\pi)$ and can be modeled approximately by
the Ornstein-Zernicke form Eq.~(\ref{eq:OZ}) around ${\mathbf Q}$. Hence the spin-fermion contribution
to the self-energy can approximately  written as (see \cite{Krien2021})
\begin{align}\label{eq:ansatz}
  \Sigma_{{\mathbf k}\, \nu= i\delta} &\propto \tilde \gamma {T} \sum_{{\mathbf q}}
\frac{ G_{({\mathbf k+ \mathbf q}) \, \nu=i \delta}} {({\mathbf q}-{\mathbf Q})^2+\xi^{-2}} \; ,
\end{align}
employing the rapid decay of  $\tilde \gamma$ with bosonic frequency $\omega$ and restricting ourselves to the Fermi energy ($\nu=i\delta$, $\delta\rightarrow 0$)\footnote{One can approximate this, at the cost of some additional broadening, by the lowest Matsubara frequency $\nu_0=i \pi T$.}. Further one can replace $G$ by $G^0$ so that Eq.~(\ref{eq:ansatz}) can also be employed as an  ansatz for the spin-fermion self-energy or for fitting the pseudogap with only two fee parameters [$\tilde \gamma$ and  $\xi$; if ${\mathbf Q}$ is known].

The sum of Eq.~(\ref{eq:ansatz})  is dominated by ${\mathbf q}\approx {\mathbf Q}$ where the antiferromagnetic spin fluctuations are strongest. In weak coupling theory   $\tilde \gamma=1$, and we get a damping, i.e., an imaginary part of the self-energy  whenever    $G_{({\mathbf k+ \mathbf Q}) 0}$ has a sizable imaginary part. This is the case whenever we have spectral weight 
$A_{({\mathbf k+ \mathbf Q}) \nu}= -\frac{1}{\pi} {\rm Im}  G_{({\mathbf k+ \mathbf Q}) \nu}$
for (real frequency) $\nu=0$. For $G=G^0$, i.e., without dampening, this is only possible if ${\mathbf k+ \mathbf Q}$  is on the Fermi surface, too. Deviations by about the  inverse correlation length (circles in Fig.~\ref{Fig:FS}) are possible, as  then  antiferromagnetic correlations
are still sizable (which is the essence of the  Ornstein-Zernicke form).

The condition that, for a considered $\mathbf k$ on the Fermi surface, also ${\mathbf k+ \mathbf Q}$ is on the Fermi surface is exactly fulfilled for the hot spots (HS) in Fig.~\ref{Fig:FS}. At weak coupling, the prefactors of Eq.~(\ref{eq:ansatz}) are small and we  need 
very long correlation lengths $\xi$ to get a sizable imaginary part of the self-energy.
Hence, here the pseudogap opens first at the hot spots.

At strong coupling the typical inverse correlation lengths are as displayed in Fig.~\ref{Fig:FS}
for the onset of the pseudogap. Numerical calculations (see~Figs.~\ref{Fig:PGSigma}, \ref{Fig:PGDGA} below) and experiment show that the pseudogap opens first at the part of the Fermi surface marked as ``PG'' in Fig.~\ref{Fig:FS}.
One reason for this might be that with the shorter correlation length, i.e., larger $\xi^{-1}$, the large spectral contribution of the  van Hove singularity 
at $(\pi,0)$\footnote{and cubic-symmetrically related momenta} becomes relevant.

A second mechanism has been identified in \cite{Krien2021}: the spin-fermion interaction $\tilde \gamma$ develops an imaginary part because of particle-hole asymmetry at strong $U$. Hence we also get an imaginary part for $\Sigma$ in Eq.~(\ref{eq:ansatz}) from the {\em real part} of the Green's function  $G_{({\mathbf k+ \mathbf q}) \, \nu =0}$:
\begin{equation}
{\rm Re}   G_{{\mathbf k} \, \nu=i\delta} = \frac{\mu -\varepsilon_{\mathbf k}-  {\rm Re} \Sigma_{{\mathbf k} \, \nu=i\delta}}{(\delta - {\rm Im} \Sigma_{({\mathbf k}) \, \nu=i\delta}  )^2+(\mu-\varepsilon_{\mathbf k}-  {\rm Re} \Sigma_{{\mathbf k} \, \nu=i\delta})^2} \; .
\end{equation}
This real part  is displayed as  false color in Fig.~\ref{Fig:FS} for the non-interacting Green's function ($\Sigma=0$).
It has opposite sign (blue vs.~red  in  Fig.~\ref{Fig:FS}) for the occupied (unoccupied) part of the Brillouin zone, where $\varepsilon_{\mathbf k}+  {\rm Re} \Sigma_{{\mathbf k} \, \nu=i\delta} <(>) \mu$; $\mu$ is the chemical potential.

Hence, for the PG momentum where we scatter into the blue occupied part in  Fig.~\ref{Fig:FS} the imaginary part of the DMFT self-energy 
is enhanced by non-local correlations. This part of the Fermi surface is hence strongly dampened and eventually develops a pseudogap.
In contrast, for the ARC momentum the imaginary part of the DMFT self-energy is even reduced (in absolute terms). The ARC quasiparticles are so-to-speak ``cooled'', i.e., become even more coherent because of spin fluctuations.
This dichotomy is another mechanism that opens the pseudogap first in the nodal (PG) region and not at the hot spot (HS) if we are at strong coupling.

 Fig.~\ref{Fig:PGSigma} shows the typical momentum differentiation that we obtain for the self-energy in the pseudogap region. The self-energy in the nodal and anti-nodal region is largely different. In the nodal direction, ${\rm Im} \Sigma_{{\mathbf k},\nu}~\sim - \nu$ with the slope corresponding to the quasiparticle weight $Z$ at this momentum. At the opening of the pseudogap, for momenta in the PG region, the self-energy first develops a large imaginary part, $\lim_{\nu\rightarrow 0}{\rm Im} \Sigma_{{\mathbf k},\nu}$. This corresponds to a strong dampening or extremely short life times of quasiparticles in this region.  In   Fig.~\ref{Fig:PGSigma} this would mean a flat curve for small frequencies.
\begin{figure}[t!]
 \vspace{-.3cm}
 
  \begin{center}
    \includegraphics[width=.6\textwidth,clip=true]{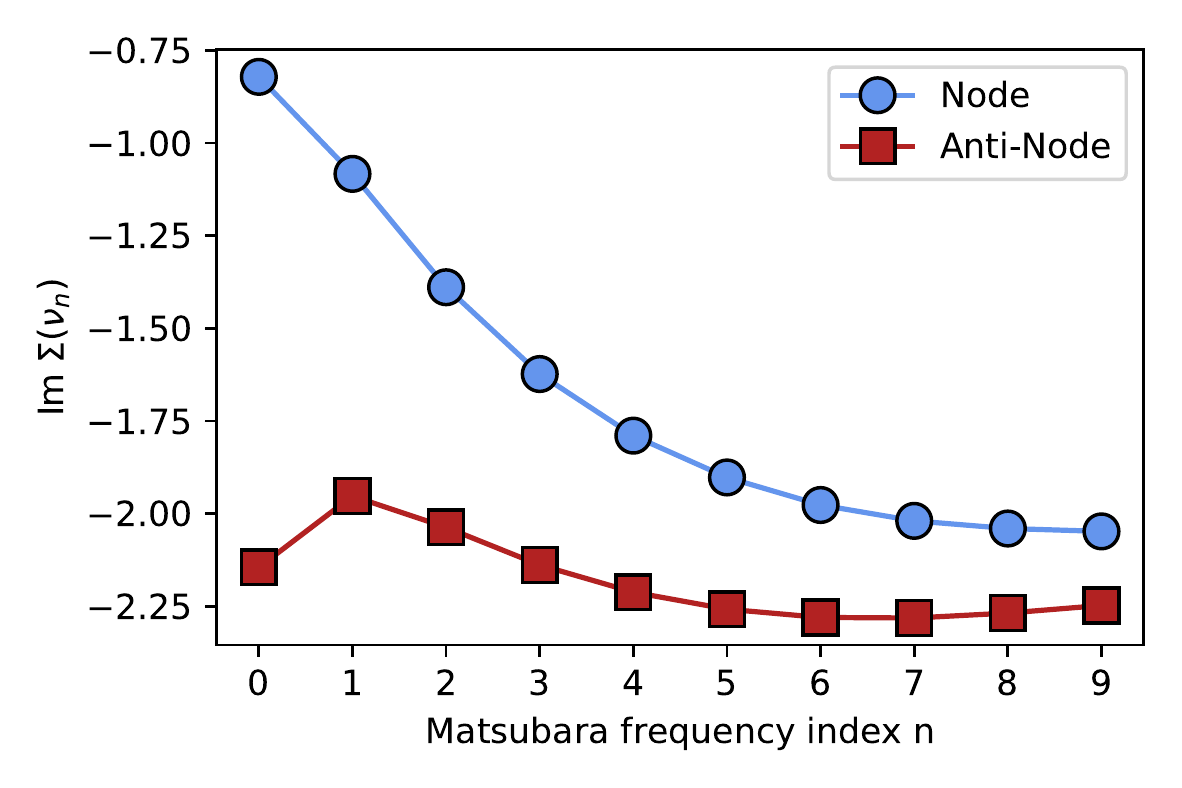}
  \end{center}
  \vspace{-.3cm}
  
  \caption{Imaginary part of the D$\Gamma$A self-energy vs.~index $n$ of the Matsubara frequency $\nu_n=i(2n+1)\pi T$, in the pseudogap region for a typical Hubbard model. Parameters: $U=8$, $t'=-0.2$, $t''=0.1$, $T=0.05$ (all in units of $t$), and 10\% hole doping. Clearly the self-energy is very different for the two momenta plotted, i..e, in the nodal direction of the Fermi surface and in the anti-nodal direction. The downturn of the anti-nodal self-energy indicates the development of a pole and thus a splitting of the spectral function away from the Fermi energy. Figure by courtesy of Paul Worm.
  \label{Fig:PGSigma}}
  \end{figure}
 Eventually, 
 the self-energy develops a pole, and we can see in   Fig.~\ref{Fig:PGSigma} the onset of such a pole as the downturn of ${\rm Im} \Sigma_{{\mathrm k},\nu}$ for small Matsubara frequencies.
 
 This is akin to the Mott insulator where the self-energy behaves as
 \begin{equation}
  \Sigma_{{\mathbf k},\nu} = \frac{U^2}{4 \nu}
 \end{equation}
 However, now this pole only occurs  in the PG momentum regime, and the origin are spin fluctuations not Mott-Hubbard physics. Consequently the prefect or, being 
 given by the spin fluctuation strength, is  smaller  than  $\frac{U^2}{4}$.
 In the PG regime, the spectrum is hence gapped because of this pole structure (or strongly suppressed if ${\rm Im}$ is only large), whereas the ARC region still shows spectral weight.

Let us now turn to the nickelate superconductors.
 Fig.~\ref{Fig:PGDGA} shows the spectrum of the Ni $3d_{x ^-y^2}$ band\footnote{There are additional Nd pockets, see Sec.~\ref{Sec:HM}.} for nickelates as calculated by D$\mathrm \Gamma$A for the effective Hubbard model (see Sec.~\ref{Sec:HM} where also the parameters can be found). The parent compound NdNiO$_2$, corresponding to $n_{3d_{x ^2-y^2}}=0.95$ (rightmost panel)  has a pseudogap in the antinodal (PG) region, indicated by missing spectral weight in  Fig.~\ref{Fig:PGDGA}. This D$\mathrm \Gamma$A prediction still awaits an experimental confirmation; angular resolved photoemission experiments are urgently needed but have not been successfully done yet. 
 
 For making nickelates superconducting, one needs to dope this
$3d_{x ^-y^2}$ band, e.g., to  $n_{3d_{x ^2-y^2}}=0.80$ or $0.85$. For these superconducting dopings,  we have a clear Fermi surface throughout the Brillouin zone in Fig.~\ref{Fig:PGDGA} and no pseudogap.\footnote{As a technical remark, the spectrum has been calculated from $A_{{\mathbf k} \nu_0}=-1/\pi \; G_{{\mathbf k} \nu_0}$ at the lowest Matsubara frequency $\nu_0= i \pi T$, which leads to some additional broadening (smearing) but avoids the error-prone and cumbersome analytical continuation.}

\section{Superconductivity}
\label{Sec:SC}
\index{superconductivity}

Next we turn to the task of calculating the superconducting order and critical temperature $T_c$
from the antiferromagnetic spin fluctuations. The general procedure has already been indicated in Fig.~\ref{Fig:ladderDGA} (c). More specifically, the first step is the calculation of the full vertex $F$ from $\Gamma_{ph}$  (plus the crossing symmetrically related $\overline{ph}$ channel), by summing  the Bethe-Salpeter ladder diagrams in this channel(s).
This calculation includes spin and charge fluctuations, but in the Hubbard model spin fluctuations prevail. In the second step, we insert these spin fluctuations  into the $pp$ channel.
That is, we calculate  $\Gamma_{pp}=\Lambda+\Phi_{ph}+\Phi_{\overline{ph}}$.
Since no $pp$ reducible diagrams have yet been included (except for the local ones), we can also rewrite this as   $\Gamma_{pp}=F-\Phi_{{pp}}^{\rm local}$. From this   $\Gamma_{pp}$ we can next recalculate $F$ including superconducting fluctuations via the Bethe-Salpeter equation in the $pp$ channel. This is akin to the Schwinger-Dyson Eq.~(\ref{eq:BSE}) in the $ph$ channel, except that we have to properly rotate the momenta and   to use $\Gamma_{pp}$ instead of   $\Gamma_{ph}$.

  Similar as in RPA [Eq.~(\ref{Eq:sus2})],  we get a superconducting (SC) susceptibility of the form
  \begin{equation}
  \chi_{{SC, \mathbf q}= {\mathbf 0}  \, \omega=0} =  \sum_{k k^\prime \nu \nu^\prime}   \underline{\underline{\chi^0}} \left[1 +  \underline{\underline{\Gamma^{\phantom{0}}_{pp,  {\mathbf q}= {\mathbf 0}  \omega=0}}} \;  \underline{\underline{\chi^0_{\phantom{q}}}}\right]^{-1}
\label{Eq:susSC}
  \end{equation}
  Here, we are interested in the instability at $\omega =0$ and the coupling of
  two fermions with momentum ${\mathbf k}$ and $-{\mathbf k}$ into a Cooper pair as in Fig.~\ref{Fig:ladderDGA} (c), i.e., a total momentum  ${\mathbf q=0}$ in the $pp$ channel\footnote{This $\mathbf q$ is not to be confused with that in the $ph$ channel which is peaked around ${\mathbf Q}=(\pi,\pi)$ and corresponds to another combination of the external legs of the four-point vertex $F$.}. The  generalized\footnote{without ${k k^\prime \nu \nu^\prime}$ summation} bare bubble susceptibility at $\omega=0$ and  ${\mathbf q=0}$ is
   \begin{equation}
     \chi^0_{{\mathbf k} {\mathbf k}^\prime \nu \nu^\prime} =G_{{\mathbf k}^\prime \nu^\prime} G_{-{\mathbf k} -\nu} \delta({\mathbf k}-\mathbf k^\prime ) \delta_{\nu, \nu^\prime}
\label{Eq:sus2chi}
  \end{equation}
   and the double underlines denote matrices with respect to  ${\mathbf k \nu}$ and $\mathbf k^\prime \nu^\prime$.

     In order to obtain superconductivity, i.e., a diverging $ \chi_{{SC, \mathbf q}= {\mathbf 0} \,  \omega=0}$ one of the eigenvalues $\lambda$  of the matrix $-\underline{\underline{\Gamma^{\phantom{0}}_{pp,  {\mathbf q}= {\mathbf 0}  \omega=0}}} \, \underline{\underline{\chi^0_{\phantom{q}}}}$
     has to approach $\lambda=1$. For electron-phonon mediated superconductivity this is simple since the electron phonon coupling gives rise to an attractive (negative) $\Gamma_{pp}$. For the repulsive Hubbard model, this is much more difficult to achieve since  $\Gamma_{pp}$ also includes the repulsive (positive) Coulomb interaction $+U$. Nonetheless, an attraction can be mediated  by antiferromagnetic spin fluctuations through retardation ($\nu, \nu^\prime$) and non-locality ($k$, $k^\prime$).
     The $k$, $k^\prime$ structure of the diverging ($\lambda\rightarrow 1$) eigenvector (or the corresponding real space structure) determines the symmetry of the superconducting symmetry breaking and gap ($s$-wave, $d$-wave etc.). For phonons with an across the board attraction, one gets $s$-wave superconductivity. For the spin fluctuations in the Hubbard model $d$-wave is more favorable to avoid the large local repulsion $+U$.\footnote{For the $d$-wave, $\Gamma_{pp}$ mediates between momenta ${\mathbf k}$ and ${\mathbf k}'$ on the Fermi surface, for which the superconducting eigenvalue has opposite sign.}
 
     The Mermin-Wagner theorem also applies for superconductivity, where in two dimensions and at finite temperatures only the Nobel-prize-winning
     topological Berezinskii–Kosterlitz–Thouless (BKT) transition is possible. At first glance, this is a contradiction to the observed 
     superconductivity in cuprates and nickelates. However, these materials are not perfectly two-dimensional but layered quasi-two-dimensional materials. In this situation, and with  the strongly increasing correlation length around the BKT transition, even a tiny coupling in the inter-layer direction will trigger superconductivity,  a little bit below the BKT transition. In the D$\mathrm \Gamma$A calculation, a self-consistency is necessary to (possibly) get a BKT transition. Without the self-consistent feedback of the superconducting fluctuations, we eventually get a mean-field kind of superconducting transition and a finite $T_c$ which is closer to experiment than the ideal two-dimensional calculation with BKT transition.

\begin{figure}[t!]
  \begin{center}
    \includegraphics[width=\textwidth,clip=true]{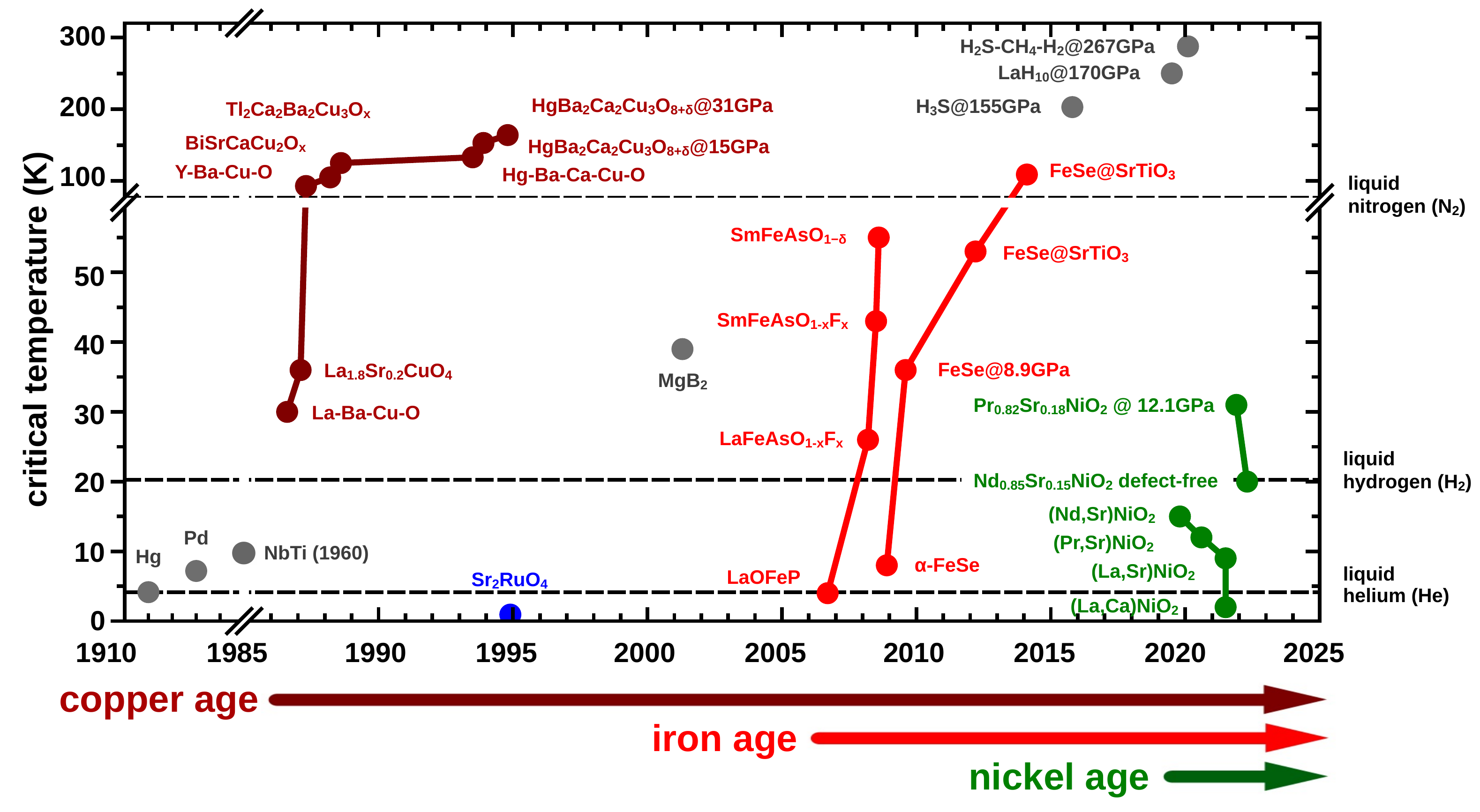}
  \end{center}
  \caption{Superconducting $T_c$ vs.\ year of discovery for various important superconductors. After the copper age, there has been an iron age, and now we are in the nickel age.  From \cite{Si2022}.  \label{Fig:NickelAge}  }
\end{figure}

     As for superconducting materials, the  three years ago  discovered nickelate superconductors \cite{li2019superconductivity} have led to enormous theoretical and experimental efforts. One therefore also speaks of the nickel age for superconductivity, see Fig.~\ref{Fig:NickelAge}.
 Also shown are hydrogen-based superconductors that are phonon-mediated and have a $T_c$ above room temperature, but only under the enormous pressure of a diamond anvil cell. Hence the challenge is to increase $T_c$ for the unconventional (i.e., not phonon-mediated) correlated superconductors or to reduce the pressure for the  hydrogen-based superconductors. Here, the nickelates have a $T_c$ that is still quite substantially below that of the cuprates. While one can expect $T_c$ to further increase somewhat with new and better synthesized nickelate films, one should not expect a room temperature nickelate superconductor. The high hope is instead that nickelates and cuprates are very similar but also decisively distinct, an ideal situation to discriminate the essentials from the incidentals for high-temperature superconductivity. The iron pnictides are instead pretty far away from the cuprate or nickelate physics.

In Sec.~\ref{Sec:HM}, we have already pointed out that the nickelates can be described by a one-band Hubbard model with an approximately adjusted doping; and in Fig.~\ref{Fig:PGDGA} we have shown the thus calculated spectrum with a pseudogap for the (non-superconducting) parent compound NdNiO$_2$.

\begin{figure}[t!]
  \begin{center}
    \includegraphics[width=.55\textwidth,clip=true]{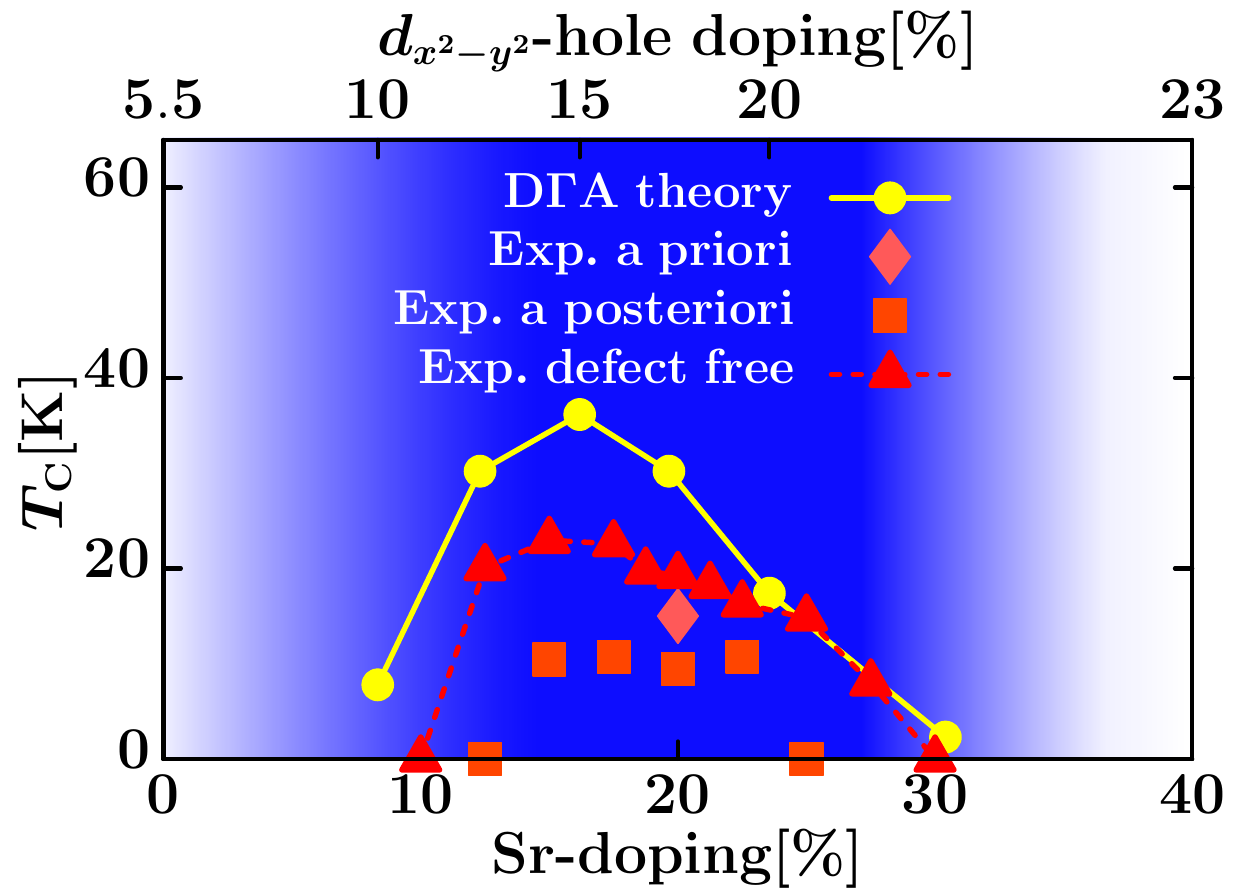}
  \end{center}
  \caption{Superconducting phase diagram $T_c$
    vs.  Sr-doping (lower $x$-axis) and  vs.~doping of the $3d_{x^2-y^2}$-orbital   (upper $x$-axis) as calculated for  Nd$_{1-x}$Sr$_x$NiO$_2$ in D$\Gamma$A \cite{Kitatani2020}. 
    At the time of the calculation a single experimental data point (``a priori'' \cite{li2019superconductivity})
    was available. The first `experimental superconducting dome (``a posteriori'' \cite{Li2020})
    showed already reasonable agreement. With cleaner films (``defect free'' \cite{Lee2022}) the agreement has become breathtaking, in particular if one considers how challenging it is to calculate $T_c$ reliably.
Also the pentalayer nickelate ($T_c\approx 13\,$K, 20\% holes in the $3d_{x^2-y^2}$ orbital\cite{Worm2021}) well matches the $T_c$ of D$\mathrm \Gamma$A and the shown 
infinite layer experiment.    Adapted from  \cite{Kitatani2020}. \label{Fig:SC}  \index{nickelates}\index{superconductivity} }
\end{figure}

For this nickelate Hubbard model, we find \cite{Kitatani2020} (not surprisingly) $d$-wave superconductivity in ladder D$\mathrm \Gamma$A. The superconducting $T_c$ vs.~doping is plotted in Fig.~\ref{Fig:SC} as a function of Sr-doping. 
Actually the theoretical calculation was a prediction here since, when three years ago superconductivity in nickelates has been discovered \cite{li2019superconductivity},  only a single $T_c$ at 20\% doping was available at first because of the difficulties to synthesize  Nd$_{1-x}$Sr$_x$NiO$_2$ in the low oxidation state Ni$^{+1}$. With the recent progress to synthesize clean superconducting nickelate films \cite{Lee2022}, the  theoretical predicted phase diagram has been spectacularly confirmed in experiment, see Fig.~\ref{Fig:SC}.

The physical reason that we see in  Fig.~\ref{Fig:SC} a superconducting dome is two-fold: The downturn at large doping is because antiferromagnetic spin-fluctuations get weaker further and further away from half-filling. The down-turn at small doping, towards half-filling, on the other hand is a consequence of the pseudogap which develops in this doping region, see Fig.~\ref{Fig:PGDGA}. Thus, the electron propagators in Fig.~\ref{Fig:ladderDGA} (c) loose coherence in a larger part of the Fermi surface, and the superconducting susceptibility is suppressed despite considerable spin fluctuations.
 
Recently also a pentalayer nickel$_{12}$ has been synthesized \cite{pan2021}, which has very 
similar hopping parameters  \cite{Worm2021}
Its $T_c=14\,$K
agrees with D$\mathrm \Gamma$A and with the $T_c$ of the infinite layer nickelates shown in Fig.~\ref{Fig:SC}. Here, DMFT indicates that pentalayer nickelates have no pockets \cite{Worm2021} and thus a  doping
of  0.2 holes per site in the $3d_{x^2-y^2}$-orbital.
The absence of pockets in pentalayer nickelates further
corroborates the picture of a decoupled reservoir that is not relevant for superconductivity as advocated in Fig.~\ref{Fig:HM}. Altogether the DFT+DMFT and      D$\mathrm \Gamma$A calculations for nickelates and the experiments for the different nickelates provide for a consistent picture \cite{Worm2021}.
This gives us some hope that we might  finally be able to actually calculate and predict superconducting $T_c$'s, the arguably biggest challenge of solid state theory.

\section{Conclusion and outlook}
\label{Sec:conclusion}
Diagrammatic extensions of DMFT are  very appealing in several ways: They combine the good description of DMFT with the, to the best of our knowledge, most important non-local physics. In this Chapter we have focused on spin fluctuations and how they mediate superconductivity, but other fluctuations such as charge fluctuations, weak localization corrections, excitons --you name it-- are treated on an equal footing. Also (quantum) criticality can be described, a topic that has been discussed in an earlier J\"ulich Autumn School \cite{Held2018}.
Diagrammatic  extensions of DMFT are also very appealing since they merry quantum field theoretical with its qualitative understanding and numerics which is unavoidable for a quantitative description of strongly correlated electrons systems.

Different variants of diagrammatic extensions of DMFT exist, and for the sake of brevity we have concentrated here on the first (and widely employed) variant: the dynamical vertex approximation.  All variants  have in common that they calculate a local vertex and
construct non-local correlations from this vertex diagrammatically.
In regions of the phase diagram where non-local correlations are short range,
results are similar as for cluster extensions of DMFT. However, the diagrammatic extensions also offer the opportunity to study long-range correlations, which  is key for (quantum) criticality but also important in other situations, as well as to calculate materials with many orbitals \cite{Galler2016}. 

There is still plenty of room for improvement: starting from (i) various self-consistencies, using (ii) a clusters instead of a single site as a starting point,  (iii) compactifying the vertex with  the intermediate representation (IR) in frequency space \cite{Wallerberger2021}, the truncated unity in momentum space
\cite{Eckhardt2020} and the single-boson exchange \cite{Krien2019} in Feynman diagram space, 
and thus allowing for parquet D$\mathrm \Gamma$A calculations at lower temperatures. Another development has been (iv) the calculation of
the underlying two-particle vertices directly for real frequencies, which is possible using the numerical renormalization group (NRG) method, see Chapter ``The physics of quantum impurity models'' by J.~von~Delft \cite{Pavarini2022}.

In this Chapter, we have concentrated on antiferromagnetic spin fluctuations, how they open a pseudogap and how they mediate superconductivity.
The $T_c$ predicted for nickelates well agrees with experiment -- actually  much better than what we dared to hoped for. This gives us some confidence that we are on the right track to better model and understand superconductivity, that we eventually have the tools  to  predict $T_c$ for new materials. While 
many theoreticians in the many-body  community consider antiferromagnetic spin fluctuations to be at the origin of high temperature superconductivity,  the mechanism for unconventional superconductivity  remains hotly debated. Maybe through a careful analysis and  predictions  we can now prove that this is indeed the microscopic mechanism for high-temperature superconductivity.

Diagrammatic extensions of DMFT such as the D$\mathrm \Gamma$A also offer the opportunity to study many other phenomena and to do material calculations. Phenomena such as changes of the topology in strongly correlated are hitherto hardly understood, a Berezinskii–Kosterlitz-Thouless transition in two-dimensions could possibly be described, or Luttinger or spin Peierls physicists in one dimension.
All of this  leaves plenty of opportunities for the next generation of physicists.

\paragraph*{Acknowledgment}
First of all, I would like  to thank my co-workers on the research presented here,  Oleg~Janson, Andrey~Katanin, Anna~Kauch, Motoharu~Kitatani,  Freidrich~Krien, Jan~M.~Tomczak, Georg~Rohringer, Thomas~Sch\"afer, Liang~Si, Alessandro~Toschi, and Paul Worm. Without them, the success of diagrammatic extensions of DMFT   would not have been possible.
Further I would like to 
thank Motoharu Kitatani and Paul Worm for carefully reading the manuscript.
Finally,  I gratefully acknowledge financially support by the Austrian Science  Fund (FWF) through project  P32044 and the  Research Unit QUAST of German Science Foundation (DFG) (DFG FOR5249; Austrian part financed through FWF project I5868).

\clearpage

\clearchapter


\end{document}